\documentclass[a4paper,12pt]{article}
\usepackage[english]{babel}
\usepackage{amsmath}
\usepackage{a4wide}
\usepackage{amsfonts}
\usepackage{graphicx}
\usepackage{color}
\usepackage{caption}
\usepackage{array}
\usepackage{pdfpages}
\usepackage{float}
\usepackage[round]{natbib}
\usepackage{multirow}
\usepackage{multicol}
\usepackage{amsxtra}
\usepackage{amsbsy}
\usepackage{accents}
\usepackage{chngcntr}
\usepackage{tabularx}
\usepackage{dcolumn}
\usepackage[none]{hyphenat}
\usepackage[affil-it]{authblk}
\usepackage{datetime}
\usepackage[a4paper]{geometry}
\usepackage[section]{placeins}

\date{\normalsize{June 2016}}
\title{\large \bf Effects of Sea Level Rise on Economy of the United States}
\author[1]{\small{Monika Novackova}}
\author[1,2,3,4,5]{Richard S.J. Tol}
\affil[1]{\small{Department of Economics, University of Sussex, Falmer, UK}}
\affil[2]{\small{Institute for Environmental Studies, Vrije Universiteit, Amsterdam, The Netherlands}}
\affil[3]{\small{Department of Spatial Economics, Vrije Universiteit, Amsterdam, The Netherlands}}
\affil[4]{\small{ Institute, Amsterdam, The Netherlands}}
\affil[5]{\small{ CESifo, Munich, Germany}}
\newcommand\starred[1]{\accentset{~~~~~\star}{#1}}
\begin{document}

\newgeometry{top=1in, bottom=1in, left=1in, right=1in}
\newdateformat{monthyeardate}{%
  \monthname[\THEMONTH], \THEYEAR}
  
\makeatletter
\def\hlinewd#1{%
\noalign{\ifnum0=`}\fi\hrule \@height #1 %
\futurelet\reserved@a\@xhline}
\makeatother

\maketitle
\vspace{3cm}

\begin{abstract}

\noindent We report the first ex post study of the economic impact of sea level rise. We apply two econometric approaches to estimate the past effects of sea level rise on the economy of the USA, viz. Barro type growth regressions adjusted for spatial patterns and a matching estimator. Unit of analysis is $3063$ counties of the USA. We fit growth regressions for $13$ time periods and we estimated numerous varieties and robustness tests for both growth regressions and matching estimator. Although there is some evidence that sea level rise has a positive effect on economic growth, in most specifications the estimated effects are insignificant. We therefore conclude that there is no stable, significant effect of sea level rise on economic growth. This finding contradicts previous ex ante studies.
\\


\end{abstract}

\pagebreak
\sloppy

\linespread{1.6}
\section{Introduction}\label{Introduction}

Sea level rise features among the more important economic impacts of climate change \citep{Tol2009}, particularly because of its potential to overwhelm regional and even national economies, either through massive land loss or exorbitantly expensive coastal protection \citep{NichollsTol}. Better understanding of past effects of sea level rise should help to predict future sea level rise effects more precisely and find optimal policies to face this consequence of climate change.

Studies of the future impact of climate change typically rely on simulation models that are applied far outside their domain of calibration \citep{IPCC5}. Model validation and parameter estimation are rare~\citep{Mendelsohn94}. This is to a degree unavoidable \textendash~ climate change is part of a yet-to-be-observed future \textendash~ but should be minimized to gain more confidence in future projections of the effects of climate change. This paper contributes by studying the economic impacts of sea level rise on the economic development of the USA in the recent past. To the best of our knowledge, no one has yet attempted to test model-based impact estimates of sea level rise against observations. This paper does not do that either. Instead, we take a key prediction from these ex ante models \textemdash that sea level rise would decelerate economic growth \textemdash and test it against the data.

Our starting point is that sea level rise is a common phenomenon. Indeed, since the start of the Holocene, global sea level rise has been 14 metres, although the bulk of it happened between seven and eight thousand years ago and most of the rest before the start of the Common Era~(\citealp{HoloceneFleming, HoloceneMilne}). Global sea level rise has been muted in more recent times \textendash~ relative to both the more distant past and future projections, but relative sea level rise has been pronounced in some locations. Thermal expansion, ice melt and ice displacement cause the sea to rise, but subsidence and tectonics can cause the land to fall \citep{IPCC5WG1}. This effect can be large. Parts of Bangkok and Tokyo, for instance, fell by five metres in a few decades during the 20\textsuperscript{th} century \citetext{\citealp{Nicholls2010, IPCC5, Tokyo2006}}.

We, however, focus on the contiguous USA, for three reasons. (\textit{i}) There are excellent data on relative sea level rise and pronounced regional differences in sea level rise. (\textit{ii}) There are also excellent data on economic growth with fine spatial detail. (\textit{iii}) Finally, regional growth patterns are well-studied in~the~USA \citetext{e.g. \citealp{Latzko2013, Higgins2006, GoetzHu}} so that we minimize the risk of ascribing to sea level rise what is caused by something else.

We hypothesize that relative sea level rise has a negative effect on economic growth. There are two main channels \textemdash see~\citet{FankTol2005} for a more thorough treatment. First, sea level rise causes damage in the form of erosion and floods, which reduce the productivity of land, labour and capital. Second, protection against coastal hazards implies that capital is diverted from productive to protective investment. On the other hand, if coastal protection is subsidized by inland areas (which may be the case in the USA), then areas with high relative sea level rise would record the economic activity of dike building etc. without suffering the costs, and would thus grow faster than other areas.

The paper proceeds as follows. Section~\ref{Method} describes the two main methods used in this study. The methods include a Barro type conditional growth regressions and a matching estimator. Section~\ref{Data} discusses data sources. Section~\ref{Empir} presents empirical results. In Section~\ref{Robust}, different variants of the Barro type economic growth regressions are discussed to verify robustness of results. Section~\ref{sConclusion} concludes.
\FloatBarrier

\section{Methodology}\label{Method}
\subsection{Barro type growth regressions}\label{GrowthM}

The rate of sea level rise changes only very slowly over time and its estimates do not vary during the relatively recent period for which economic data are available. Therefore, we opted for cross-sectional regressions rather than panel data analysis. Conventional growth regressions are fitted according to~\citet{Barro91} and~\citet{Barro92}. As a staring point, average growth rate of per capita income is regressed on the initial logarithm of per capita income and on sea level rise without other covariates. After that, other covariates are added that have been found to be important in previous studies. The regression equation can be written as:

\begin{equation}\label{growth}
g_n = \alpha + \beta y_{n,0} + \gamma 'x_n + v_n,
\end{equation}

where $y_{n,0}$ is the initial logarithm of per capita income in county $n$, $g_n = (y_{n,T}-y_{n,0})/T$ is average growth rate of per capita income between years $0$ and $T$ for county $n$, $y_{n,T}$ is the logarithm of per capita income in year $T$, $x_n$ is a vector of controls capturing regional differences and $v_n$ is an error term which is assumed to have zero mean and finite variance. The controls in~$x_n$ are listed in~Table~\ref{tVariables} in~Appendix~1 and discussed below. Coefficient $\beta$ is typically found to be negative, that is, poorer regions grow faster than richer.

\citet{Evans97} shows that the  OLS estimator of~\eqref{growth} is consistent and unbiased only if the following conditions are satisfied: (\textit{i}) The dynamical structures of economies can be expressed by identical AR(1) processes; (\textit{ii}) every economy affects every other economy symmetrically; and (\textit{iii}) all permanent cross-economy differences are captured by control variables. As these conditions are highly implausible, \citet{Evans97} suggested a three stage least square (3SLS) estimation method to obtain consistent estimates. In the first and second stage, the following equation is estimated using an IV estimator: 

\begin{equation}\label{GrowtDiff}
\Delta g_n = \omega + \beta \Delta y_{n,0} + \eta_n,
\end{equation}

where $\Delta$ denotes first difference. Thus, the first stage involves the estimation of:

\begin{equation}\label{FirstSt}
\Delta y_{n,0} = \delta 'z_n + \xi_n,
\end{equation}

where $z_{n}$ is a vector of instruments, $\delta$ is a vector of parameters to be estimated and $\xi_n$ is the error term.
The predicted values of $ \Delta y_{n,0}$ from~\eqref{FirstSt} are used to estimate the second stage:

\begin{equation}\label{FirstSec}
\Delta g_n = \kappa + \beta \hat{\delta} 'z_n + \zeta_n,
\end{equation}

where $\hat{\delta}$ are the OLS estimates of $\delta$ from~\eqref{FirstSt} so that $\hat{\delta} 'z_n$ is the predicted value of $ \Delta y_{n,0}$ from~\eqref{GrowtDiff}.
Then the variable $\pi_n=g_n - \hat{\beta} y_{n,0}$ is created using the estimate $\hat{\beta}$ from~\eqref{FirstSec} and in the third stage the following regression is estimated:

\begin{equation}\label{Third}
\pi_n = \tau + \starred{\gamma} 'x_n + \epsilon_n,
\end{equation}

where~$\tau$ and~$\starred{\gamma}$ are parameters and~$\epsilon_n$ is the error term. 

The model estimated in this paper explains economic growth during the period 1990\nobreakdash-2012, thus year~zero is 1990 and $T = 22$. As in~\citet{Higgins2006}, asymptotic conditional convergence rates are calculated by substituting estimate of $\beta$ from equation~\eqref{FirstSec} into the formula~$c=1-(1+T\beta)^{1/T}$. Estimates of~$\starred{\gamma}$ from~\eqref{Third} represent initial effects on economic growth rate rather than partial effects on average growth rate. However, if~$\beta$ is negative \textendash~ as assumed by the neoclassical growth hypothesis \textendash~ the signs of these estimates will be the same as the signs of partial effects of the elements in~$x_n$ on average economic growth rate. Also, under the assumption that~$\beta$ is identical across the counties, the magnitude of the coefficients relative to one another is the same as the magnitude of the partial effects of the variables in~$x_n$ relative to one another.

Matrix $x_n$ includes the control variables that are important to achieve conditional convergence. If they were not included, the model would represent the hypothesis of absolute convergence rather than the hypothesis of conditional or club convergence~\citep{Higgins2006}. It was found by previous literature (\citealp{Rupasingha, GoetzHu}) that these covariates have an effect on economic growth \textendash~ hence they can affect the relationship between growth and sea level rise if correlated with sea level rise. Furthermore, the inclusion of control variables reduces the risk of omitted variables bias and the standard errors of estimates are smaller. 

An important covariate is distance from coast as the absolute value of its correlation coefficient with sea level rise is extremely high compared to other covariates, because sea level rise is zero for all inland counties. The value of the correlation coefficient is $-0.336$ and its \textit{p}-value is lower than $2.2 \times 10^{-16}$. Furthermore, the coastal counties are different because of their transport facilities and natural amenities. Other important covariates are per capita highway and education expenditures and per capita tax income, which accounts for total taxes imposed by local government. The highway and education expenditures are included as a measure of local government expenditure and the tax income is a measure of local government activities. These controls are relevant, because they are related to decisions about funding of dikes and other forms of coastal protection. Besides, it is believed that higher taxes tend to deter potential immigrants and discourage people from starting a business which may slow down economic growth. On the other hand, higher government infrastructure expenditure might attract entrepreneurs.

The other covariates are sorted into four groups, particularly measures of agglomeration, measures of religious adherence, regional dummy variables and other socioeconomic and environmental indicators.

The measures of religious adherence are included since \citet{Rupasingha} show that religious adherence has significant impact on economic growth. Moreover, the included religious variables are correlated with a dummy variable which indicates presence of interstate highways. Therefore, these variables are relevant to our study as dike building is usually funded from the same sources as the construction of highways. More details about included covariates can be found in Table~\ref{tVariables} in~Appendix~1. Descriptive statistics of these variables are summarized in Tables~\ref{tDescriptive} and~\ref{tDescriptiveA} in Appendix~2.

The instruments in $z_n$ in equations~\eqref{FirstSt} and~\eqref{FirstSec} are chosen from the set of $1980$ values of the explanatory variables with the exemption of interstate highway access, state right to work laws, amenity scale, regional and rural/urban indicator variables. The criterion for the choice of instruments was the Sargan test of overidentifying restrictions. It turned out that the test is insignificant when per capita religious adherence and population density are used as instruments. These two covariates are therefore used in $z_n$ in~\eqref{FirstSt} and~\eqref{FirstSec}. Although the Sargan test is not considered as a very strong criterion, it is clear that all possible instruments are exogenous as they are from year 1980 and the dependent variable is economic growth for the period starting in year 1990. In order to confirm the appropriateness of the IV~estimation we used the Wu-Hausman test which is described for example in~\citet{Mackinn}. The value of the test statistic is $9.502$ and the corresponding \textit{p}-value is $0.002$, thus the null hypothesis of exogenity is rejected, which is in accordance with the growth model estimation theory presented by~\citet{Evans97}. 

As the analysis is based on cross county data, we may expect the data to be spatially dependent. According to~\citet{LeSageIntro}, spatial dependence in the dependent variable causes OLS estimates to be biased and spatial dependence in error terms causes OLS estimates to be inefficient. To obtain unbiased and efficient estimates an approach which takes the spatial dependency into account is needed. 

As in~\citet{LeSage98}, the general spatial model for~\eqref{Third} can be written as follows:
\begin{equation}\label{genSpat}
\begin{array}{lcll}
\pi &~=~& \rho W \pi +X \beta + u, \\
u &~=~& \lambda W u + \epsilon, \\
\epsilon &~\sim ~& N(0,\sigma^2I_n), \\
\end{array}
\end{equation}

where $\pi$ is a~$n \times 1$ vector of dependent variables, scalar~$\rho$ is a spatial lag parameter, scalar~$\lambda$ is a spatial error parameter,~$W$ is the known $n \times n$ spatial weight matrix, $X$ is an $n \times k$ matrix of explanatory variables that determine the growth, $\beta$ is $k \times 1$ vector of parameters and $\epsilon$ is the error term.

In this study, the binary contiguity matrix $W$ is constructed as a symmetric matrix where $W_{ij}=1$ if county $i$ and county $j$ have a common border and $W_{ij}=0$ otherwise. Since it is unrealistic to assume that no spillover effects exist between island counties and counties which are close to them, the island counties are treated as if they had common borders with coastal counties which surround them. Matrix~$W$ is row standardised, which means that the sum of all $W_{ij}$ is equal to~$n$. 

Model~\eqref{genSpat} considers two spatially autoregressive processes, in particular a spatial process in the dependent variable and a spatial process in error terms. Imposing restrictions on~\eqref{genSpat}, more specific spatial models can be derived. Setting~$\rho=0$ produces a spatial error model, which can be written as in~\citet{LeSage98}: 

\begin{equation}\label{SERM}
\begin{array}{lcll}
\pi &~=~& X \beta + u, \\
u &~=~& \lambda W u + \epsilon, \\
\epsilon &~\sim ~& N(0,\sigma^2I_n). \\
\end{array}
\end{equation}

Imposing restriction $\lambda=0$ on equations~\eqref{genSpat} results in a spatial autoregressive model~(SAR). According to~\citet{LeSage98} this model can be written as:

\begin{equation}\label{SAR}
\begin{array}{lcll}
\pi &~=~& \rho W \pi +X \beta + \epsilon, \\
\epsilon &~\sim ~& N(0,\sigma^2I_n). \\
\end{array}
\end{equation}
As is shown in Section~\ref{Empir}, specification~\eqref{SAR} is the most appropriate, therefore we estimate this specification and use it as the basis for further variations and robustness tests. The model is estimated via maximum likelihood estimation. First the parameter $\rho$ is found applying a one dimensional optimization procedure; $\beta$ and the other parameters are subsequently found by generalized least squares.

Models~\eqref{SAR} were estimated for various time periods to verify whether the results remain the same. In particular, $13$ models with~$T$ from $10$ to $22$ were estimated and these are discussed in Section~\ref{Empir}. Year zero is 1990 in all of these models. Matrix $X$ in~\eqref{SAR} contains the same set of covariates for all~$13$ models. Each covariate in these $13$ models is from the same year (which is stated in Table~\ref{tVariables} in Appendix~1 for individual covariates).

\subsection{Matching estimator}\label{MatchingM}

Matching is a technique used to estimate the effect of a treatment
\citetext{see \citealp{MatchingKniha} and \citealp{Matching}}. In this study we use it to verify our results obtained by the Barro type growth regressions. An advantage of matching is that a functional form does not need to be specified, thus it is not susceptible to misspecification bias. Furthermore, as only matched cases are used, less weight is put on outliers.

The treatment effect estimator, which assumes that suitable matching has already been found, is described in the next few paragraphs. After that a procedure of creating a suitable matching and its assessment is discussed.

Let $y_0$ denote the outcome of interest without treatment, $y_1$ the outcome of interest with treatment and $d$ a dummy variable which is equal to $1$ for treated and $0$ for untreated individuals. As shown in~\citet{MatchingKniha}, if $E(y_0|d,X)=E(y_0|X)$ the mean treatment effect on the treated $E(y_1 - y_0|d=1) $ is identified with $E\{y-E(y|X,d=0)|d=1 \} $. The estimator used in this study can be written as:

\begin{equation}\label{TreatmentEffect}
T_N \equiv N^{-1}_u \sum_{i \in T_u} ( y_i - |C_i|^{-1} \sum_{m \in C_i} y_{mi} ), \\
\end{equation}
where $N_u$ is the number of successfully matched treated subjects, $T_u$ is the set of the successfully matched treated subjects, $y_i$ is a response variable in treated $i$, $C_i$ is a group of controls assigned to treated $i$, $|C_i|$ is a number of controls in comparison group $C_i$ and $y_{mi}$ denotes a response variable in $C_i$. The standard errors are estimated according to~\citet{AbadieImSE}.

Instead of matching on $X$, one may get around the dimensionality problem by matching on one dimensional propensity score $\pi(X)$ for which it holds $\pi(X) \equiv P(d=1|X)$. The propensity score is the probability for an individual to participate in a treatment given his observed covariates~$X$. It is shown in~\citet{MatchingKniha} that if~$d$ is independent of~$(y_0,y_1)$ given~$X$, it is also independent of~$(y_0,y_1)$ given just~$\pi(X)$.

To estimate a propensity score, we have to choose a model to be estimated and a set of variables to be included in the model. We fitted several types of models, including a binomial logistic regression (logit), a probit and a linear probability model. According to quality of matching, the~most suitable is logistic regression and probit. The models are fitted by iteratively reweighted least squares.

The literature suggests several ways to select explanatory variables for the propensity score (see e.g.~\citealp{MatchingKniha, Matching}). Here, the variables are chosen according to their statistical significance and according to quality of matching.

Matchings obtained by different methods are evaluated and compared according to measures of imbalance. The main emphasis is put on the \textit{p}-values of two sided \textit{t}\nobreakdash-tests of equality of means of the successfully matched treated and successfully matched controls and on \textit{p}-values of Kolmogorov-Smirnov tests of the null hypothesis that the probability density of the successfully matched treated is the same as density of successfully matched controls. The test statistics are calculated for each variable in $X$ separately.

In this case, the treatment is sea level rise and the variables to be matched on are the covariates from model~\eqref{SAR} listed in Table~\ref{t3SLSSAR}. We considered all inland counties and four counties with negative sea level rise as controls. Since the sea level rise is not a binary variable, we decided to consider all coastal counties with difference of the sea level rise and its $95\%$ confidence interval higher than a~certain value as treated. The $95\%$ confidence intervals were obtained from the same source as the mean sea level trends and they are inversely related to length of sea level data collection period. The data sources are discussed in Section~\ref{Data}. As the length of confidence intervals is independent of sea level rise and economic growth, the use of confidence intervals to define the set of treated should not cause the matching estimator to be biased.

Since the dataset contains only $274$ coastal counties, which is much less than the number of controls, we chose the threshold for defining the treated observations to be equal to a ten percent sample quantile of sea level rise of coastal counties, which is $1.8$ mm/year.
\footnote{We also tried other matching algorithms besides propensity score matching. These include Mahalanobis distance and its generalization, where the optimal weights of each covariate are found by a generic search algorithm~\citep{GenMatch}. In this case, the best matchings in terms of balance are obtained applying the propensity score method, therefore results of other matchings are not presented.}

\FloatBarrier
\section{Data}\label{Data}

All control variables used in this study are listed in Table~\ref{tDescriptive} or Table~\ref{tVariables} in~Appendix~1. Since values of some of these covariates are not available for all counties, most of the models are estimated using a~dataset which includes~$3063$ counties for which all data are available, while the total sample size is~$3072$. Descriptive statistics of sea level rise, average growth rate of per capita income and the most relevant covariates are summarized in Table~\ref{tDescriptive}. Descriptive statistics of the other covariates can be found in Table~\ref{tDescriptiveA} in~Appendix~2. The statistics are calculated for the sample of complete cases. 

\begin{table}[H] \centering
\caption{\textbf{Descriptive Statistics }}\label{tDescriptive}
\centering
\begin{tabular}{lrr}
\hlinewd{1pt}
\textbf{Variable} & \textbf{Mean} & \textbf{Std. dev.} \\ 
\hlinewd{1pt}
Sea level rise - stations average (mm/year)& $ 2.764 $ & $ 1.768 $\\
Sea level rise - coastal counties (mm/year) &  $3.376$ &$2.068$ \\
\hline
Average growth rate of per capita income 1990-2012 & \multirow{2}{*}{$ 0.041 $} & \multirow{2}{*}{$0.008 $} \\
\hspace{0.3cm} (Income in log of dollars)  & & \\
\hline
Coast distance (km) & $600.914$ & $463.532$ \\
\hline
Gov. expenditures per capita (Thousands of US\$) & $ 1071.411$ & $ 376.838 $ \\
Tax income per capita (Thousands of US\$) & $ 652.926 $ & $ 434.457 $ \\
\hline
\end{tabular}
\end{table}
\pagebreak

The sea level rise data are available at the website of the~\citet{NOAA} (CO-OPS). The water level data were collected at $94$ CO-OPS water gauge stations located within the contiguous United States. Water levels have been captured at these stations for a span of at least 30 years. The fact that the sea level data collection period varies across the water gauge stations may make the analysis more complicated. This issue is addressed in Section~\ref{Range}. According to information provided by CO-OPS, the sea level trends were obtained by the decomposition of the sea level variations into a linear secular trend, an average seasonal cycle, and residual variability at each station. Estimated $95\%$ confidence intervals of the sea level trends are also available at the website of CO\nobreakdash-OPS. For most of the stations, water level data~up~to~the~year~2007 were used for estimation of mean sea level trend.

The sample of complete data includes $ 274 $ coastal counties and $ 2789 $ inland ones. The $94$ CO-OPS stations are located in $86$ coastal counties.
We considered the sea level rise to be equal to zero in the inland counties. For the coastal counties extrapolation is needed. A simple extrapolation is adopted as follows. For a few coastal counties with more than one station, the sea level rise is calculated as the arithmetic average of the sea level trend captured at different stations in county. For counties with one CO-OPS station, the mean sea level trend measured at this station is used. For counties with no CO-OPS station, the sea level rise is obtained as mean sea level trend, measured at the station which is closest to the centroid of the county. The distance is calculated as the shortest Euclidean distance. The $95\%$ confidence intervals of sea level rise are extrapolated in the same way as mean sea level trends. In Section~\ref{Avgextrapolation}, we apply a different extrapolation as a robustness test.

Since most of the counties are landlocked with zero sea level rise, it makes little sense to present descriptive statistics of sea level rise of the whole sample. Therefore Table~\ref{tDescriptive} shows the mean and standard deviation of sea level rise for the sample of $94$ CO-OPS stations and the mean and standard deviation of sea level rise of the subsample of coastal counties using the extrapolation described above. 

The per capita income growth data for all years are drawn from the Bureau of Economic Analysis.  Descriptive statistics of per capita income growth rates for the $13$ relevant time periods are summarized in Table~\ref{tGrowthDes} in~Appendix~3. Distance from coast was obtained as the shortest Euclidean distance from centroids of counties to coast. Details about the data sources of the other covariates can be found in Appendix~2.

\FloatBarrier
\section{Empirical results}\label{Empir}

In Section~\ref{GrowthEmpi}, the empirical results of several variants of Barro type growth models are presented. The empirical results of the matching estimator discussed in Section~\ref{MatchingM} are presented in Section~\ref{MatchEmpi}.

\subsection{Barro type growth regressions}\label{GrowthEmpi}

As a starting point, we fitted a single OLS regression of economic growth $g_n$ on sea level rise without any other covariates and an OLS regression of economic growth $g_n$ on sea level rise and its square without any other covariates. Estimates of these two regressions and estimates of a 3SLS model characterised by equations~\eqref{GrowtDiff}~to~\eqref{Third} without other covariates are summarized in Table~\ref{tNoCov}. 

We also included sea level rise squared. If the squared term is not included, the linear term will be positive and slightly significant in some of the models. This is not in accordance with our expectation and the reason may be the nonlinearity of the relationship. Therefore, the quadratic term of sea level rise is included and it turns out to be negative in most cases and often significant.

\begin{table} \centering 
\caption{} 
\label{tNoCov} 
\begin{tabular}{@{\extracolsep{-1pt}}lccc} 
\\[-1.8ex]\hline 
\vspace{-0.65cm}\\
\hlinewd{2pt}\\[-1.8ex] 
 & OLS 1 & OLS 2 & 3SLS \\ 
 
 \vspace{-1cm} \\
  & &  & equation~\eqref{Third}\\

 \vspace{-1cm} \\
 
\\[-1.8ex]Dependent variable & $g$ & $g$ & $\pi$ \\ 
\hlinewd{2.2pt}
Constant & 0.077 (0.011)$^{***}$ & 0.077 (0.011)$^{***}$ & $-$1.390 (0.008)$^{***}$ \\ 
\hline
Log of initial per capita & \multirow{2}{*}{$-$0.004 (0.001)$^{***}$} & \multirow{2}{*}{ $-$0.004 (0.001)$^{***}$ } & \multirow{2}{*}{0.146 (0.036)$^{***}$ } \vspace{-0.2cm}\\ 
\hspace{0.3cm}  income (US\$) &  & &  \\
\hline
Sea level rise (m/year) & 0.828 (0.145)$^{***}$ & 0.565 (0.497) & $-$4.077 (3.875) \\ 
\hline
Sea level rise (m/year) -  & \multirow{2}{*}{$---$ }& \multirow{2}{*}{26.340 (47.610)} & \multirow{2}{*}{901.900 (367.900)$^{*}$} \vspace{-0.2cm}\\ 
\hspace{0.3cm}  squared &  &  & \\ 
\hlinewd{1.5pt}
\textbf{Measures \vspace{-0.2cm}} &\multirow{2}{*}{\textbf{No}} &\multirow{2}{*}{\textbf{No}} & \multirow{2}{*}{\textbf{No}}\\
\textbf{\hspace{0.3cm} of agglomeration} & & & \\
\hline
\textbf{Measures \vspace{-0.2cm}} &\multirow{2}{*}{\textbf{No}} &\multirow{2}{*}{\textbf{No}} &\multirow{2}{*}{\textbf{No}}\\
\hspace{0.3cm} \textbf{of religious adherence} & & &\\
\hline
\textbf{Other socioeconomic \vspace{-0.2cm}} &\multirow{3}{*}{\textbf{No}} &\multirow{3}{*}{\textbf{No}} &\multirow{3}{*}{\textbf{No}}\\
\hspace{0.3cm} \textbf{and environmental \vspace{-0.2cm}} & & & \\
\hspace{0.3cm} \textbf{indicators} & & & \\
\hline
\textbf{Regional dummy \vspace{-0.2cm}} & \multirow{2}{*}{\textbf{No}} &\multirow{2}{*}{\textbf{No}}&\multirow{2}{*}{\textbf{No}} \\
\textbf{ \hspace{0.3cm} variables} & & & \\
\hlinewd{1.5pt}
Convergence rate & 0.004 & 0.004 & 0.004 \\ 
\hlinewd{1.5pt}
Observations & 274 & 274 & 274 \\ 
\\[-1.8ex]\hline 
\vspace{-0.95cm}\\
\hlinewd{2pt}\\[-1.8ex] 

\textit{Notes:} & \multicolumn{3}{r}{Standard errors in brackets} \\ 
 & \multicolumn{3}{r}{$^{*}$p$<$0.05; $^{**}$p$<$0.01; $^{***}$p$<$0.001} \\ 
\end{tabular} 
\end{table} 

In the first column of Table~\ref{tNoCov}, the effect of sea level rise is positive and significant, whereas the literature has assumed the opposite effect. However, as mentioned above, the OLS estimate of Barro type growth regression is not consistent in most cases. Furthermore the possible relationship between sea level rise and economic growth can be non-linear. The peculiar result may also be due to omitted variable bias. When the squared sea level rise is included, both linear and square terms are positive and insignificant. Things change for the 3SLS estimate. Income diverges, as the log of initial per capita income in the third column is positive. The linear term of sea level rise is negative and insignificant, while the quadratic term is positive and slightly significant. These results might be biased as other covariates are omitted and spatial patterns are not taken into account, therefore more accurate models are estimated.

OLS estimates of model~\eqref{growth} for period 1990-2012 with covariates can be found in Table~\ref{tOLS} in~Appendix~3. The 3SLS estimates of equation~\eqref{Third} for the same period including covariates can be found in the first column of Table~\ref{t3SLSSAR}. Adjusted R-squared is $0.492$ for this model and value of \textit{F}-statistic is~$119.8$ with a~\textit{p}-value lower than $2.2 \times 10^{-16}$. Estimates of the first stage~\eqref{FirstSt} and the second stage~\eqref{FirstSec} of this model are summarized in Table~\ref{tStage12} in~Appendix~3. 
However, as possible spatial relationships are not taken into account, these estimates may be biased and inconsistent.

Moran's I confirms spatial dependence for the economic growth rate~$g_n$. The test statistic equals~$0.500$ with a \textit{p}-value lower than~$2.2 \times 10^{-16}$, thus the null hypothesis of no spatial dependence is rejected. Moran's I was calculated also for the variable~$\pi_n$ from equation~\eqref{Third}. Its value is~$0.532$ and the corresponding \textit{p}-value is lower than~$2.2 \times 10^{-16}$. Also in this case, the null hypothesis of no spatial dependence is rejected. One of the forms~\eqref{genSpat},~\eqref{SERM} or~\eqref{SAR} should therefore be fitted instead of applying the usual 3SLS procedure. 

As an additional check whether the use of the spatially adjusted model is justified, we used the Lagrange Multiplier~(LM)~diagnostic tests for spatial dependence as proposed by~\citet{Bera96}. Specifically, we used the LM test for spatial error dependence and the LM test for a~missing spatially lagged dependent variable. We also calculated variants of these tests, which are robust to presence of the other. These include the LM test for spatial error dependence in the presence of omitted spatially lagged dependent variable and the other way around. Distributions of these test statistics are well known for the case of OLS residuals, therefore we applied them to residuals from~\eqref{growth} and to residuals from~\eqref{Third}. The values of the LM statistics for spatial error dependence and for missing spatially lagged dependent variable and its robust versions \citep{Bera96} are summarized in Table~\ref{LM}.

\vspace{1.5cm}
\begin{table}[H] \centering
\caption{\textbf{LM tests for spatial dependence in residuals}}\label{LM}
\vspace*{0.5cm}
\begin{tabular}{llcc|cc|}
\cline{3-6}
\multicolumn{2}{c}{} & \multicolumn{2}{|c}{\multirow{2}{*}{ }} & \multicolumn{2}{|l|}{\textbf{Missing }} \\
\vspace{-1.1cm} \\
\multicolumn{2}{c}{} & \multicolumn{2}{|c}{\textbf{Error dependence}} & \multicolumn{2}{|l|}{\textbf{spatially lagged}} \\ 
\vspace{-1.1cm} \\
\multicolumn{2}{c}{} & \multicolumn{2}{|c}{}& \multicolumn{2}{|l|}{\textbf{dependent variable}} \\ 
\cline{3-6}
\multicolumn{2}{c|}{} & \bf{Test} & \multirow{2}{*}{\bf{\textit{p}-value}}&\bf{Test } & \multirow{2}{*}{\bf{\textit{p}-value}} \\ 
\vspace{-1.1cm} \\
\multicolumn{2}{c|}{} & \bf{statistic} & &\bf{statistic} &  \\ 
\hline
\multicolumn{1}{|c}{\textbf{OLS \eqref{growth}}} &\multicolumn{1}{|c|}{ \textbf{Standard}} & $625.270$ & $<2.2 \times 10^{-16}~$& $631.655$ & $<2.2 \times 10^{-16}~$\\
\multicolumn{1}{|c}{\textbf{residuals }} & \multicolumn{1}{|c|}{\textbf{Robust}} & $22.527$ & $2.072 \times 10^{-6}~$& $28.912$&$7.575 \times 10^{-8}~$ \\
\hline 
\multicolumn{1}{|c}{\textbf{3SLS \eqref{Third}}} & \multicolumn{1}{|c|}{\textbf{Standard}} & $553.635$ & $<2.2 \times 10^{-16}~$ & $533.797$ & $<2.2 \times 10^{-16}~$ \\
\multicolumn{1}{|c}{\textbf{residuals }} & \multicolumn{1}{|c|}{\textbf{Robust}} & $41.802$ & $1.010 \times 10^{-10}~$ & $21.964$& $2.779 \times 10^{-6}~$ \\
\hline
\end{tabular}
\end{table}
\vspace{1.5cm}

All statistics in Table~\ref{LM} are highly significant, suggesting that a~general spatial model~\eqref{genSpat} could be a~suitable form. Estimates of this form are summarized in the first column of Table~\ref{tGenWhite} in Appendix~3. Parameter~$\lambda$ is insignificant while~$\rho$ is highly significant which indicates that specification~\eqref{SAR} is more suitable. Estimates~of~\eqref{SAR} are summarized in the second column of Table~\ref{t3SLSSAR}, the estimates of all coefficients including the covariates can be found in the second column of Table~\ref{t3SLSSARall} in Appendix~3. Also according to the LM~test for residual autocorrelation, specification~\eqref{SAR} is appropriate. The value of this test statistic is $0.826$ and its \textit{p}-value is $0.364$, thus the null hypothesis of uncorrelated error terms is not rejected. Model~\eqref{SAR} is therefore taken as a starting point for further analysis and for estimation of different variants of this model.

\begin{table}\centering 
\caption{} 
\label{t3SLSSAR} 
\begin{tabular}{@{\extracolsep{-1pt}}lcc} 
\\[-1.8ex]\hline 
\hline \\[-1.8ex] 
 \multicolumn{3}{r}{\textbf{Income growth model for period 1990-2012}} \\ 
\cline{2-3}
& 3SLS  & SAR  \\ 
\vspace{-1cm} \\
& model~\eqref{Third} &  model~\eqref{SAR} \\ 
\hline \\[-1.8ex] 
Constant & 0.348 (0.002)$^{***}$ & 0.185 (0.007)$^{***}$ \\
Log of initial per capita income (US\$)& $-$0.033 (0.005)$^{***}$ & $-$0.033 (0.005)$^{***}$ \\ 
Sea level rise (m/year)& 0.947 (0.277)$^{***}$ & 0.594 (0.252)$^{*}$ \\ 
Sea level rise (m/year) - squared & $-$59.200 (37.040) & $-$44.406 (33.711) \\ 

Coast distance (thousands km) & $-$0.007 (0.001)$^{***}$ & $-$0.005 (0.001)$^{***}$ \\ 
Coast distance (thousands km) - squared & 0.008 (0.001)$^{***}$ & 4,535.100 (690.000)$^{***}$ \\ 
Gov. expenditures per capita (billion US\$)& $-$0.710 (0.451) & $-$0.596 (0.411) \\ 
Tax income per capita (billion US\$)& 4.171 (0.399)$^{***}$ & 3.370 (0.368)$^{***}$ \\ 
$\rho$ (SAR) & --- & 0.458 (0.021)$^{***}$ \\ 
\hline
\textbf{Measures of agglomeration} &\textbf{Yes} &\textbf{Yes} \\
\textbf{Measures of religious adherence} &\textbf{Yes} &\textbf{Yes} \\
\textbf{Other socioeconomic } &\multirow{2}{*}{\textbf{Yes}} &\multirow{2}{*}{\textbf{Yes}}\\
\hspace{0.3cm} \textbf{and environmental indicators} & &\\
\textbf{Regional dummy variables} &\textbf{Yes} &\textbf{Yes} \\
\hline \\[-1.8ex] 
Convergence rate & 0.058 & 0.058  \\ 
\hline \\[-1.8ex] 
Observations & 3,063 & 3,063 \\ 
\hline 
\hline \\[-1.8ex] 
\textit{Notes:} & \multicolumn{2}{r}{Standard errors in brackets} \\ 
& \multicolumn{2}{r}{$^{*}$p$<$0.05; $^{**}$p$<$0.01; $^{***}$p$<$0.001} \\ 
\end{tabular} 
\end{table}

As we can see in the second column of Table~\ref{t3SLSSAR}, the sea level rise is positive and slightly significant, while the squared sea level rise is negative and insignificant in spatial autoregressive model~\eqref{SAR}.

 As explained in~\citet{LeSageIntro}, impact measures are needed for correct interpretation of coefficients of models with spatially lagged dependent variable. Because of the spillover effects, a change in explanatory variable in one observation can potentially effect value of dependent variable of all other observations. Therefore the coefficients can not be interpreted in the same way as typical OLS coefficients. 
 
 The impact measures for our model~\eqref{SAR}, which are summarized in Table~\ref{tSARImpacts}, were calculated according to equation~$2.46$~(\citealp{LeSageIntro}, p. $38$) using exact dense matrix. A direct impact is an impact of an explanatory variable in county~\textit{i} on the dependent variable in county~\textit{i}, indirect impact is an impact of an explanatory variable in county~\textit{i} on the dependent variable in all counties but~\textit{i} and total impact is~a sum of~direct and indirect impact. The impacts of all covariates included in this model can be found in Table~\ref{tSARImpacts2} in Appendix~3.
 
 \pagebreak
\begin{table}[H] \centering 
\caption{} 
\label{tSARImpacts} 
\begin{tabular}{@{\extracolsep{-1pt}}lrrr} 
\hline 
\hline \\[-1.8ex] 
\multicolumn{4}{c}{\textbf{Income growth model for period 1990-2012 - Impact measures}} \\ 
\hline
\\[-1.8ex] & \multicolumn{3}{c}{ SAR model~\eqref{SAR}} \\ 
\\[-1.8ex] &  Direct & Indirect & Total\\ 
\hline \\[-1.8ex] 
Sea level rise (m/year) &  0.6218  & 0.4753 & 1.0971 \\ 
Sea level rise (m/year) - squared & $-$46.4611 & $-$35.5122 &  $-$81.9733\\ 
Coast distance (thousands km) & $-$0.0048 & $-$0.0036 & $-$0.0084 \\ 
Coast distance (thousands km) - squared & 4,744.9020 & 3,626.7320 & 8,371.6340 \\ 
Gov. expenditures per capita (billion US\$) & $-$0.6232 & $-$0.4764 & $-$1.0996 \\ 
Tax income per capita (billion US\$) & 3.5257 & 2.6948 & 6.2205\\  
\hline
\textbf{Measures of agglomeration} &\multicolumn{2}{r}{\textbf{Yes}} \\
\textbf{Measures of religious adherence} &\multicolumn{2}{r}{\textbf{Yes}}\\
\textbf{Other socioeconomic } &\multicolumn{2}{r}{\multirow{2}{*}{\textbf{Yes}}}\\
\vspace{-1cm} \\
\hspace{0.3cm} \textbf{and environmental indicators} &\\
\textbf{Regional dummy variables} &\multicolumn{2}{r}{\textbf{Yes}} \\
\hline 
\hline \\[-1.8ex] 
\end{tabular} 
\end{table} 
\pagebreak

\begin{figure}[H] 
\includegraphics[width=16.7cm]{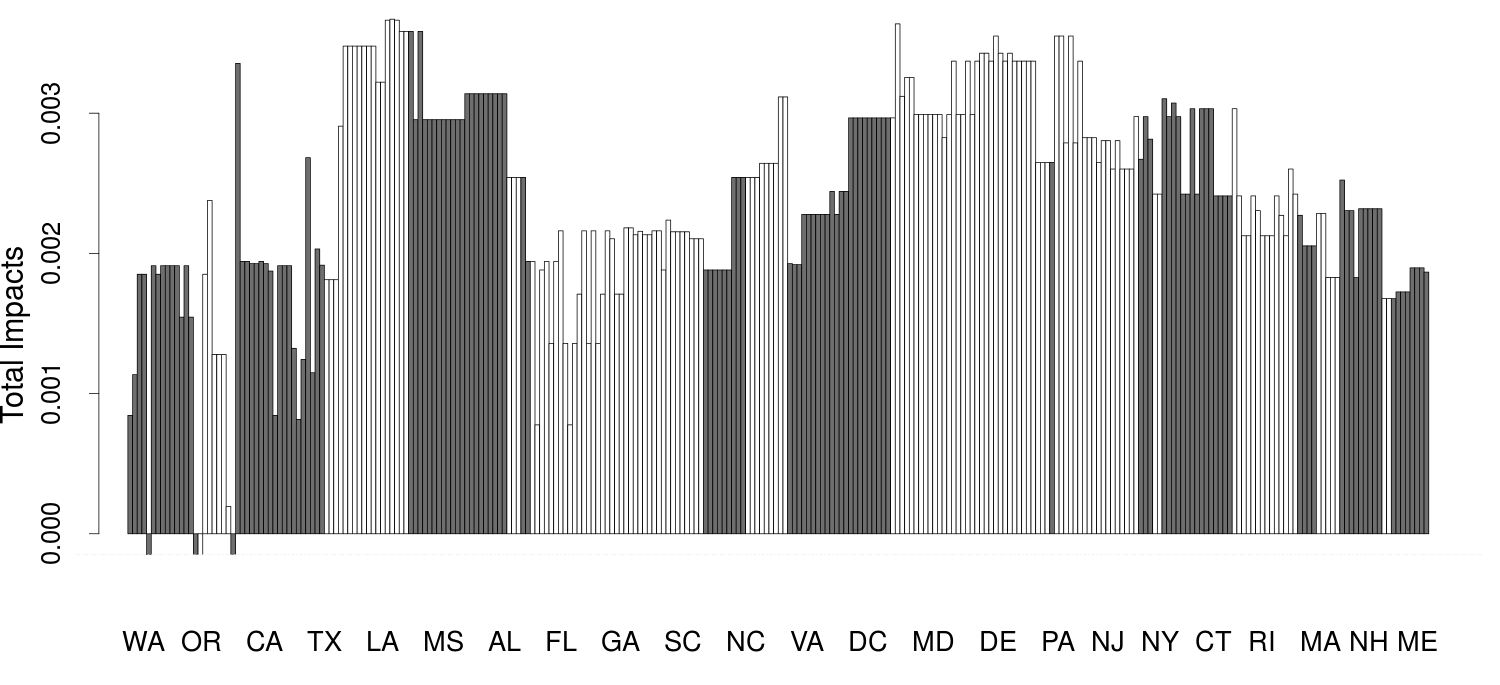}
\caption{Initial effects of sea level rise on economic growth rate - Total Impacts}\label{bar}
\end{figure}

The coefficients in Table \ref{t3SLSSAR} are barely significant but we show effect size nonetheless. Estimated total initial impacts of sea level rise on the economies of coastal counties of United States are depicted in Figure~\ref{bar}. We obtained the counties' impacts by multiplying the sea level rise and its square of each county with the estimated total impacts of sea level rise (which can be found in the first two rows of~Table~\ref{tSARImpacts}). In~Figure~\ref{bar}, the counties are ordered according to their location along the coast, first west coast from north to south, then the counties along the Gulf of Mexico and after that east coast from south to north. The alternating black and white groups of bars represent groups of counties in each coastal state. Perhaps surprisingly given the parameters, the impacts are only negative in the four counties where sea level is falling.

As mentioned above, we estimated model~\eqref{SAR} for different time periods of economic growth. In~total we estimated $13$ different models for $13$ different time periods, which are listed in the first column of~Table~\ref{tPeriods1}. The first row relates to time period 1990\nobreakdash-2012, hence this row depicts the same estimates of sea level rise and coast distance as those that can be found in the second column of~Table~\ref{t3SLSSAR}. 

\begin{table} \centering
\caption{\textbf{Sea level rise and coast distance estimates:}}\label{tPeriods1} 
\caption*{SAR models~\eqref{SAR} for different time periods}
\begin{tabular}{|l|cc|cc|cc|cc|}
\hline
 & \multicolumn{4}{c|}{\textbf{SLR }}  & \multicolumn{4}{c|}{\textbf{Coast}} \\

\multicolumn{1}{|c|}{\textbf{ Period}} & \multicolumn{4}{c|}{\textbf{ }} &  \multicolumn{4}{c|}{\textbf{distance}}\\
\cline{2-9}
&\multicolumn{2}{l|}{\textbf{Linear}}& \multicolumn{2}{c|}{\textbf{Squared}} &\multicolumn{2}{l|}{\textbf{Linear}} & \multicolumn{2}{c|}{\textbf{Squared}} \\
\hline
$1990-2012$& + & * & - & & - & *** & + & *** \\
$1990-2011$ & + & & +& &- &&+&*** \\
$1990-2010$ & + & $\bullet$ & -& & - & ** &+&***\\
$1990-2009$ & + & *** & - & ** & - & ** & + & *** \\
$1990-2008$ & - & & + & &- & $ \bullet $ & +& **\\
$1990-2007$ & - & & + & & - & & + & * \\
$1990-2006$ & + & *** & - & ** & - & $\bullet$ & + & *\\
$1990-2005$ & +& *** & - & *** & - & * & + & *** \\
$1990-2004$ & + & *** & - & *** & - & * & + & *** \\
$1990-2003$ & + & & - & & - & ** & + & *** \\
$1990-2002$ & + & *** & - & *** & - &* & + & ** \\
$1990-2001$ & + & *** & - & *** & - & ** & + & ***\\
$1990-2000$ & + & *** & - & *** & - & ** & + & ***\\
\hline
\multicolumn{1}{|l|}{Observations:} & \multicolumn{8}{r|}{3063} \\ 
\hline
\multicolumn{1}{l}{\textit{Notes:}} & \multicolumn{8}{r}{All models include all covariates from~Table~\ref{t3SLSSARall}} \\ 
\multicolumn{1}{l}{ } & \multicolumn{8}{r}{$+$ estimate is positive; $-$ estimate is negative} \\ 
\multicolumn{1}{l}{} & \multicolumn{8}{r}{$^{\bullet}$p$<$0.1;$^{*}$p$<$0.05; $^{**}$p$<$0.01; $^{***}$p$<$0.001} \\ 
\end{tabular}
\end{table}

As one can see in Table~\ref{tPeriods1}, for the period 1990\nobreakdash-2006 and the shorter periods both linear and quadratic sea level rise terms are significant and the linear term is positive while the quadratic term is negative. The period 1990\nobreakdash-2003 is the exception: sea level rise is insignificant. However, for most of the longer periods both linear and quadratic sea level rise terms are insignificant, therefore it can not be generally claimed that sea level rise has a significant effect on economic growth. The relationship between sea level rise and economic growth is unstable over time. As~the growth rates are averaged over the periods in~Table~\ref{tPeriods1}, we see that the relationship reverses in 2003,~2007 and~2011. The only interpretation is therefore that the earlier significance is a fluke.

\FloatBarrier
\subsection{Matching estimator}\label{MatchEmpi}

We compared a number of different propensity score matchings.  Methods used to obtain these matchings differ in variables in balance matrix, caliper, number of controls assigned to one treated, propensity score model, whether the matching is with replacement or not and in way how ties are treated.  Specifically, we found three different matchings with balance achieved on all covariates listed in Table~\ref{t3SLSSARall} except for sea level rise and coast distance. We excluded coast distance from the balance matrix as all treated counties are coastal, while most of the controls are inland, thus it would be impossible to obtain matching balanced on this variable. For the three balanced matchings, two sided \textit{t}-tests of equality of means and both naive and bootstrap Kolmogorov-Smirnov tests are insignificant for all the covariates. All these three matchings are paired matchings with one control assigned to each treated and without replacement. Ties are randomly broken.

The estimated treatment effect and some features of the three completely balanced matchings are summarized in Table~\ref{tMatching}. The explanatory variables in each propensity score model estimated in this study are covariates of the corresponding balance matrix. Regarding the first matching in Table~\ref{tMatching}, the balance matrix and the propensity score model include all covariates listed in Table~\ref{t3SLSSARall} with the exception of sea level rise and coast distance. It also includes the square of government expenditures, nonwhites, and amenities. The propensity score model of the second and the third matching in Table~\ref{tMatching} includes also squared percentage of Catholics besides the explanatory variables included in the propensity score model for the first matching.

\vspace{1.5cm}
\begin{table}[H] \centering
\caption{\textbf{Balanced propensity score matchings}}\label{tMatching}
\begin{center}
\begin{tabular}{crccccc}
\hline
&
\multicolumn{1}{c}{\textbf{Estimated}} &\textbf{Std.} &\textbf{\textit{p-}} &\textbf{Treated} & \textbf{Prop.} & \\
\vspace{-1.15cm} \\
\textbf{Matching} &
\multicolumn{1}{c}{\textbf{treatment}} &\textbf{error}& \textbf{ value} &\textbf{matched} & \textbf{score } &\textbf{Caliper} \\
\vspace{-1.15cm} \\
& \multicolumn{1}{c}{\textbf{effect}} && &\textbf{cases} & \textbf{model} & \\
\hline 
$1$& $8.60\times 10^{-5}$ & $2.12\times 10^{-4}$ & $0.684$ & $131$ & Logit &$0.035$ \\
$2$& $-6.46\times 10^{-5}$ & $1.85\times 10^{-4}$ & $0.726$ & $136$ & Probit &$0.035$ \\
$3$& $1.88\times 10^{-5}$ & $1.89\times 10^{-4}$ & $0.921$ & $126$ & Probit &$0.020$ \\
\hline
\multicolumn{1}{l}{\textit{Notes:}} & \multicolumn{6}{r}{Estimated effect: Treatment effect for the treated} \\ 
\multicolumn{1}{l}{ } & \multicolumn{6}{r}{Caliper in multiples of standard deviation for each covariate} \\ 
\end{tabular}
\end{center}
\end{table} 
\vspace{1.5cm}

The estimated treatment effect for the treated is positive for the first and third matching, and negative for the second matching. In all three cases the effect is insignificant. Besides these three matchings we estimated a number of other matchings, however balance was not achieved on all relevant covariates for them. For almost none of these not completely balanced matchings, the estimate of the treatment effect is significantly different from zero. As in the case of the economic growth model, no significant effect of sea level rise on economy of the United States was found applying the matching estimator.

\FloatBarrier

\section{Robustness}\label{Robust}
Variants of the models discussed in Section~\ref{GrowthEmpi} are estimated to test the robustness of our findings.

\subsection{Heteroscedasticity}\label{White}

We estimated heteroscedasticity robust White estimates to find out whether the model does not suffer from more general types of heteroscedasticity. Specifically, we fitted the following spatial lag model: 

\begin{equation}\label{WhiteEq}
\begin{array}{lcll}
\pi &~=~& \rho W \pi +X \beta + \epsilon. \\
\end{array}
\end{equation}
The model was estimated by performing a generalized two stage least square procedure~\citep{Stsls} with a heteroscedasticity correction to the covariances of coefficients to obtain a~White consistent estimator. We used the spatially lagged values of variables in $X$ as instruments for the spatially lagged dependent variable. The White estimates are compared with the estimates of the spatial autoregressive lag model~\eqref{SAR} in Table~\ref{tWhite}. They do not differ substantially. The full set of estimates can be found in the second column of Table~\ref{tGenWhite} in Appendix~3.

The impact measures for model~\eqref{WhiteEq} calculated according to equation~$2.46$~(\citealp{LeSageIntro}, p. $38$) using exact dense matrix can be found in Table~\ref{tWhiteImpacts} in Appendix~3.

\begin{table} \centering 
\caption{} 
\label{tWhite} 
\small
\begin{tabular}{@{\extracolsep{-1pt}}lcc} 
\\[-1.8ex]\hline 
\hline \\[-1.8ex] 
 \multicolumn{3}{r}{\textbf{Income growth model for period 1990-2012}} \\ 
\cline{2-3} \\[-1.8ex] 
& SAR model~\eqref{SAR} & White errors~\eqref{WhiteEq} \\ 
\hline \\[-1.8ex] 
Constant & 0.185 (0.007)$^{***}$ & 0.177 (0.019)$^{***}$ \\ 
Log of initial per capita income (US\$)& $-$0.033 (0.005)$^{***}$ & $-$0.033 (0.005)$^{***}$ \\ 
Sea level rise (m/year)&  0.594 (0.252)$^{*}$ & 0.577 (0.244)$^{*}$ \\ 
Sea level rise (m/year) - squared & $-$44.406 (33.711) & $-$43.675 (31.879) \\ 
Coast distance  (thousands km) & $-$0.005 (0.001)$^{***}$ & $-$0.004 (0.001)$^{***}$ \\
Coast distance - squared & \multirow{2}{*}{ 4,535.100 (690.000) $^{***}$ }&\multirow{2}{*}{  4,347.300 (844.850)$^{***}$} 
\vspace{-1cm} \\ 
\\ 
\hspace{0.3cm} (thousands km squared) & &  \\ 

Gov. expenditures per capita & \multirow{2}{*}{ $-$0.596 (0.411)} & \multirow{2}{*}{$-$0.590 (0.570) }\\ 
\vspace{-1cm} \\ 
\hspace{0.3cm} (billion US\$) & &  \\ 
Tax income per capita (billion US\$) & 3.370 (0.368)$^{***}$ & 3.330 (0.543)$^{***}$ \\ 
$\rho$ (SAR) & 0.458 (0.021)$^{***}$ & 0.481 (0.054)$^{***}$ \\ 
 \hline
\textbf{Measures of agglomeration} &\textbf{Yes} &\textbf{Yes} \\
\textbf{Measures of religious adherence} &\textbf{Yes} &\textbf{Yes} \\
\textbf{Other socioeconomic } &\multirow{2}{*}{\textbf{Yes}} &\multirow{2}{*}{\textbf{Yes}}\\
\vspace{-1cm} \\ 
\hspace{0.3cm} \textbf{and environmental indicators} & &\\
\textbf{Regional dummy variables} &\textbf{Yes} &\textbf{Yes} \\
\hline \\[-1.8ex] 
Convergence rate & 0.004 & 0.004 \\ 
\hline \\[-1.8ex] 
Observations & 3,063 & 3,063 \\ 
\hline 
\hline \\[-1.8ex] 
\textit{Notes:} & \multicolumn{2}{r}{Standard errors in brackets} \\ 
 \multicolumn{3}{r}{$^{*}$p$<$0.05; $^{**}$p$<$0.01; $^{***}$p$<$0.001} \\ 
\end{tabular} 
\end{table} 
\FloatBarrier

\subsection{Outliers}

The spatial autoregressive models were estimated without outliers for all $13$ periods. We considered all observations with sea level rise or average growth rate of per capita income higher or equal to its 95th sample percentile or lower or equal to its 5th sample percentile as outliers. Estimates of sea level rise and coast distance coefficients of the models without outliers are compared with estimates of the models based on the whole sample in Table~\ref{tOOutliers}. Columns~$(2) - (5)$ include estimates of the models for the whole sample and estimates of the models without outliers are presented in columns~$(6) - (9)$. The sea level rise coefficients estimated using the sample without outliers are insignificant in all $13$ models except one. This confirms the conclusion that sea level rise has no significant effect on economic growth as it seems that the previously significant results were mostly driven by outliers.

\begin{table} \centering
\caption{\textbf{Sea level rise and coast distance estimates:}}\label{tOOutliers} 
\caption*{SAR models~\eqref{SAR} for different time periods}
\begin{small}
\begin{tabular}{|l|cc|cc|cc|cc|cc|cc|cc|cc|}
\hline
& \multicolumn{8}{c|}{\textbf{Whole sample}} & \multicolumn{8}{c|}{\textbf{Without outliers}} \\
\cline{2-17}
\multicolumn{1}{|c|}{\textbf{ Period}} & \multicolumn{4}{c|}{\textbf{SLR }}  & \multicolumn{4}{c|}{\textbf{Coast}} & \multicolumn{4}{c|}{\textbf{SLR}}  & \multicolumn{4}{c|}{\textbf{Coast}} \\

& \multicolumn{4}{c|}{\textbf{ }} &  \multicolumn{4}{c|}{\textbf{distance}} & \multicolumn{4}{c|}{} &  \multicolumn{4}{c|}{\textbf{distance}}\\
\cline{2-17}

&\multicolumn{2}{l|}{\textbf{Linear}}& \multicolumn{2}{c|}{\textbf{Sq.}} &\multicolumn{2}{l|}{\textbf{Linear}} & \multicolumn{2}{c|}{\textbf{Sq.}} &\multicolumn{2}{c|}{\textbf{ Linear}} & \multicolumn{2}{c|}{\textbf{Sq. }} & \multicolumn{2}{c|}{\textbf{ Linear}} & \multicolumn{2}{c|}{\textbf{Sq.}} \\
\hline 
$1990-2012$& + & * & - & & - & *** & + & *** & -& & + & &-& * & + & *** \\
$1990-2011$ & + & & +& &- &&+&** &+ & & - & & - & & + & *\\
$1990-2010$ & + & $\bullet$ & -& & - & ** &+&*** & + & & - & & - & & + & ***\\
$1990-2009$ & + & *** & - & ** & - & ** & + & *** &  - & & + & $\bullet$ & - & ** & + & ***\\
$1990-2008$ & - & & + & &- & $ \bullet $ & +& ** & + & & - & & - & & + & $\bullet$\\
$1990-2007$ & - & & + & & - & & + & * & + & & - & & + & & + & \\
$1990-2006$ & + & *** & - & ** & - & $\bullet$ & + & * & - & & + & * & - & ** & + & ***\\
$1990-2005$ & +& *** & - & *** & - & * & + & *** & - & & + & $\bullet$ & - & ** & + & ***\\
$1990-2004$ & + & *** & - & *** & - & * & + & *** & - & & + & & - & * & + & ***\\
$1990-2003$ & + & & - & & - & ** & + & *** & + & & - & & - & & + & * \\
$1990-2002$ & + & *** & - & *** & - &* & + & ** & - & & + & $\bullet$ & - & *** & + & *** \\
$1990-2001$ & + & *** & - & *** & - & ** & + & *** & + & & - & & - & * & +& *** \\
$1990-2000$ & + & *** & - & *** & - & ** & + & *** & - & & + & & - & *** & + & ***\\
\hline
\multicolumn{1}{|l|}{Obs.:} & \multicolumn{8}{r|}{$3063$} & \multicolumn{8}{r|}{ Varies between $2593$ and $2607$} \\ 
\hline
\multicolumn{1}{l}{\textit{Notes:}} & \multicolumn{16}{r}{All models include all covariates from~Table~\ref{t3SLSSARall}} \\ 
\multicolumn{1}{l}{} & \multicolumn{16}{r}{$+$ estimate is positive; $-$ estimate is negative} \\ 
\multicolumn{1}{l}{} & \multicolumn{16}{r}{$^{\bullet}$p$<$0.1;$^{*}$p$<$0.05; $^{**}$p$<$0.01; $^{***}$p$<$0.001} \\ 
\end{tabular}
\end{small}
\end{table} 

All models in Table~\ref{tOOutliers} include the covariates listed in Table~\ref{t3SLSSARall}, but the estimates are not presented here to save space. The signs and significance levels of the coast distance coefficients are depicted as they are highly correlated with sea level rise. 

\FloatBarrier

\subsection{Groundwater depletion}\label{Depletion}
One reason why no significant negative effect was found can be a reverse causality due to groundwater depletion. An alternative hypothesis is that excessive ground water withdrawal has led to land subsidence which appears as relative sea level rise. More water is being extracted in more populated areas with higher economic growth, thus higher economic growth can be positively correlated with relative sea level rise, which may cancel the negative effects of sea level rise on the economy. 

Groundwater depletion has only been an issue in some coastal areas in United States~\citep{Depletion}. As~a~robustness test we estimated the spatial autoregressive models (for the $13$ time periods) for subsamples without the coastal areas that experience groundwater depletion. The estimates of~\citet{Depletion} were used to sort the states where groundwater has been depleted into four groups according to volume of depleted water during the relevant time period. Then, the model was estimated for four subsamples. First the model was estimated for the subsample without the states in the group with the highest levels of depletion, then for the subsample without the two groups with the highest levels of depletion, after that the three groups of states with the highest levels of depletion were excluded and finally all four groups were excluded. For the subsample without the first group, the estimates of sea level rise coefficients do not differ significantly from the complete sample for almost all time periods. For the other three subsamples, previously significant sea level rise coefficients are not significant any more, which can be also due to decreased sample size. These results are in accordance with the above conclusion that no significant effect of sea level rise was detected.

\FloatBarrier

\subsection{Sea level data sample range}\label{Range}

The period of sea level data collection varies across the CO-OPS stations. Since the length of~data collection period is independent of sea level rise or economic growth, it should not cause a~measurement error or bias. However, the unequal length of collection periods may cause a~heteroscedasticity problem. The possible heteroscedasticity issue is discussed in Section~\ref{White} and as one can see in Table~\ref{tWhite}, the heteroscedasticity robust White estimates do not differ substantially from the estimates of~\eqref{SAR} thus heteroscedasticiy is not a substantial issue.

As a further robustness test, we fitted the models for all 13 time periods of economic growth using the mean sea level trend estimated for identical $28$ years long time periods using water level data available at the website of~\citet{PSMSL} (PSMSL). The maximum length of time period for which the data are available for most of the stations is $28$ years, specifically from the year 1979 until 2007. These data are only available for water gauge stations in $57$ counties, thus we used extrapolated values of sea level rise for the other counties. The same way of extrapolation is applied as described in Section~\ref{Data}. In~Table~\ref{t30yrs}, the signs and significance levels of coefficients obtained by our basic variant of~\eqref{SAR} (using the whole sea level rise data collection periods) are compared with the estimates obtained using the $28$ years long time period of sea level rise data collection. The table summarises $13$ models for the $13$ time periods of economic growth, each row corresponds to one time period. Although these models include also all other covariates from Table~\ref{t3SLSSARall}, only the sea level rise and coast distance coefficients are presented in~Table~\ref{t30yrs} to save space. The results do not differ substantially, significance levels and signs of the sea level rise are the same for most of the time periods.

\begin{table} \centering
\begin{small}
\caption{\textbf{Sea level rise and coast distance estimates:}}\label{t30yrs} 
\caption*{SAR models~\eqref{SAR} for different time periods}
\begin{tabular}{|l|cc|cc|cc|cc|cc|cc|cc|cc|}
\hline
& \multicolumn{8}{c|}{\textbf{Full range of SLR data}} & \multicolumn{8}{c|}{\textbf{ SLR data from $1979-2007$}} \\
\cline{2-17}
\multicolumn{1}{|c|}{\textbf{ Period}} & \multicolumn{4}{c|}{\textbf{SLR }}  & \multicolumn{4}{c|}{\textbf{Coast}} & \multicolumn{4}{c|}{\textbf{SLR}}  & \multicolumn{4}{c|}{\textbf{Coast}} \\

& \multicolumn{4}{c|}{\textbf{ }} &  \multicolumn{4}{c|}{\textbf{distance}} & \multicolumn{4}{c|}{} &  \multicolumn{4}{c|}{\textbf{distance}}\\
\cline{2-17}

&\multicolumn{2}{l|}{\textbf{Linear}}& \multicolumn{2}{c|}{\textbf{Sq.}} &\multicolumn{2}{l|}{\textbf{Linear}} & \multicolumn{2}{c|}{\textbf{Sq.}} &\multicolumn{2}{c|}{\textbf{ Linear}} & \multicolumn{2}{c|}{\textbf{Sq. }} & \multicolumn{2}{c|}{\textbf{ Linear}} & \multicolumn{2}{c|}{\textbf{Sq.}} \\
\hline 
$1990-2012$& + & * & - & & - & *** & + & *** & + & $ \bullet $ & - & & - & *** &+& *** \\
$1990-2011$ & + & & +& &- &&+&*** &-& & + & & -& & + & *** \\
$1990-2010$ & + & $\bullet$ & -& & - & ** &+&*** & + & & - & & - & *** & + & *** \\
$1990-2009$ & + & *** & - & ** & - & ** & + & *** & + & *** &- &* &- & *** & + & *** \\
$1990-2008$ & - & & + & &- & $ \bullet $ & +& ** & - & & + & & - & $ \bullet $ &+& *** \\
$1990-2007$ & - & & + & & - & & + & * & - & & + & & - & & + & *\\
$1990-2006$ & + & *** & - & ** & - & $\bullet$ & + & *  & + & *** & -& ** &-& ** & + &**\\
$1990-2005$ & +& *** & - & *** & - & * & + & *** & + & *** &- & *** & - & ** & + & ***\\
$1990-2004$ & + & *** & - & *** & - & * & + & *** & + & *** & - & *** & - & ** & + & ***\\
$1990-2003$ & + & & - & & - & ** & + & *** & + & & - & & - & ** & + & *** \\
$1990-2002$ & + & *** & - & *** & - &* & + & ** & + & *** & -& *** & - & ** & + & *** \\
$1990-2001$ & + & *** & - & *** & - & ** & + & *** & + & *** & - & ** & - & *** & + & *** \\
$1990-2000$ & + & *** & - & *** & - & ** & + & *** & + & ** & - & * & - & *** & + & ***\\
\hline
\multicolumn{1}{|c|}{Obs.:} & \multicolumn{8}{r|}{$3063$} & \multicolumn{8}{r|}{ $3063$} \\ 
\hline
\multicolumn{1}{l}{\textit{Notes:}} & \multicolumn{16}{r}{All models include all covariates from~Table~\ref{t3SLSSARall}} \\ 
\multicolumn{1}{l}{} & \multicolumn{16}{r}{$+$ estimate is positive; $-$ estimate is negative} \\ 
\multicolumn{1}{l}{} & \multicolumn{16}{r}{$^{\bullet}$p$<$0.1;$^{*}$p$<$0.05; $^{**}$p$<$0.01; $^{***}$p$<$0.001} \\ 
\end{tabular}
\end{small}
\end{table}

All coefficients of the two models in the first row of~Table~\ref{t30yrs} are compared in~Table~\ref{t30YrCompare} in~Appendix~3. Thus, Table~\ref{t30YrCompare} compares estimates of~\eqref{SAR} using the sea level rise data from the whole data collection ranges (our basic specification summarised in the second column of~Table~\ref{t3SLSSAR}) with estimates of the same specification using sea level rise data from the shortened~$28$~years long time period. In both of~these models the time period of economic growth is $1990$-$2012$. We can see that the estimates and their significance levels are very similar in these two specifications. Regarding the models for the other $12$~periods of economic growth in~Table~\ref{t30yrs}, estimates of other coefficients not presented in~Table~\ref{t30yrs} are also very similar to estimates obtained using the whole ranges of sea level rise data collection. However, they are not presented here to save space.

We can conclude that the results are robust with respect to time period of the sea level rise data collection.

\FloatBarrier

\subsection{Sea level rise extrapolation}\label{Avgextrapolation}

Since not every coastal county has a CO-OPS water gauge station, the sea level rise variable was extrapolated. As a test of robustness, models were fitted using another method of extrapolation. For coastal counties without CO-OPS station, sea level rise was calculated as the average of the sea level trend over all $94$ CO-OPS stations weighted by inverse Euclidean distance between each station and centroid of the county. The sea level rise of counties with at least one CO-OPS station was obtained in the same manner as above.

The results do not differ substantially from those above. The signs of estimates and the significance levels are the same for most of the covariates for both extrapolations for all $13$ time periods. For the linear sea level rise term, there is no change in sign or significance level for any time period. The effect of squared sea level rise term changes from significant to insignificant in one case (period 1990-2009) when using the weighted average way of extrapolation. The results are reasonably robust with respect to method of extrapolation of sea level rise.

\FloatBarrier

\subsection{Coastal and near coast counties}

According to Pearson's product-moment correlation coefficient, sea level rise and distance from coast are significantly correlated. The value of the test statistic is~$-0.335$ and the corresponding~\textit{p}-value is lower than $2.2\times 10^{-16}$. Because this may cause one of these coefficients to~capture the effect of the other, spatial autoregressive models~\eqref{SAR} with all covariates are re-estimated for the subsample of counties which are near the coast and for the subsample of~coastal counties. Another reason why comparison of models for these subsamples with models for all counties can be revealing, is the fact that sea level rise only directly affects the coastal counties.

Models estimated using the whole sample are compared with the models estimated for the subsample of counties which are near the coast in Table~\ref{tPeriods}. Columns~$(2) - (5)$ include estimates of the models using the whole sample, therefore they are the same as those in Table~\ref{tPeriods1}. Columns~$(6) - (9)$ in Table~\ref{tPeriods} describe models estimated for the subsample of counties which are near the coast. These counties were defined based on the shortest Euclidean distance between coast and centroid of each county. The subsample of near coast counties includes $761$ counties for which the distance between centroid and coast is shorter than $189$km, which is the first quartile
of the sample distribution of the shortest distances between counties' centroids and the coast.

\begin{table} \centering
\begin{small}
\caption{\textbf{Sea level rise and coast distance estimates:}}\label{tPeriods} 
\caption*{SAR models~\eqref{SAR} for different time periods}
\begin{tabular}{|c|cc|cc|cc|cc|cc|cc|cc|cc|}
\hline
& \multicolumn{8}{c|}{\textbf{All counties}} & \multicolumn{8}{c|}{\textbf{ Near coast counties}} \\
\cline{2-17}
\textbf{Period } & \multicolumn{4}{c|}{\textbf{SLR }}  & \multicolumn{4}{c|}{\textbf{Coast}} & \multicolumn{4}{c|}{\textbf{SLR}}  & \multicolumn{4}{c|}{\textbf{Coast}} \\

\textbf{ } & \multicolumn{4}{c|}{\textbf{ }} &  \multicolumn{4}{c|}{\textbf{distance}} & \multicolumn{4}{c|}{} &  \multicolumn{4}{c|}{\textbf{distance}}\\
\cline{2-17}
&\multicolumn{2}{l|}{\textbf{Linear}}& \multicolumn{2}{c|}{\textbf{Sq.}} &\multicolumn{2}{l|}{\textbf{Linear}} & \multicolumn{2}{c|}{\textbf{Sq.}} &\multicolumn{2}{c|}{\textbf{ Linear}} & \multicolumn{2}{c|}{\textbf{Sq.}} & \multicolumn{2}{c|}{\textbf{ Linear}} & \multicolumn{2}{c|}{\textbf{Sq.}} \\
\hline 
$1990-2012$& + & * & - & & - & *** & + & *** & + & & + & & - & $\bullet$ & + & $\bullet$ \\
$1990-2011$ & + & & +& &- &&+&*** & +& & -& &+ & &- & \\
$1990-2010$ & + & $\bullet$ & -& & - & ** &+&***&+ & &+ & & -& & +& \\
$1990-2009$ & + & *** & - & ** & - & ** & + & *** & -& &+ & &- & ** &+ & * \\
$1990-2008$ & - & & + & &- & $ \bullet $ & +& ** &+ & &+ & &+ & &- & \\
$1990-2007$ & - & & + & & - & & + & * &+ & &- & &+ & &- & \\
$1990-2006$ & + & *** & - & ** & - & $\bullet$ & + & * &- & &+ & &- & *&+ &* \\
$1990-2005$ & +& *** & - & *** & - & * & + & *** & + & &- & &- & *& +& *\\
$1990-2004$ & + & *** & - & *** & - & * & + & *** & +& &- & &- &* &+ &* \\
$1990-2003$ & + & & - & & - & ** & + & *** & +& * &- & * &- & &+ & \\
$1990-2002$ & + & *** & - & *** & - &* & + & ** & +& &- & &- &* &+ &* \\
$1990-2001$ & + & *** & - & *** & - & ** & + & *** & + & &- & &- &* &+ &$\bullet$ \\
$1990-2000$ & + & *** & - & *** & - & ** & + & *** & + & &- & &- & *& +& *\\
\hline
\multicolumn{1}{|c|}{Obs.:} & \multicolumn{8}{r|}{$3063$} & \multicolumn{8}{r|}{ $761$} \\ 
\hline
\multicolumn{17}{l}{\textit{Notes:} All models include all covariates from~Table~\ref{t3SLSSARall} except of dummy variables \vspace{-0.2cm}} \\ 
\multicolumn{17}{l}{ \hspace{1.2cm} for the following regions: Great Lakes, Plains, Southwest and Rocky Mountain, \vspace{-0.2cm}} \\ 
\multicolumn{17}{l}{ \hspace{1.2cm} which are not included in the models for the coastal counties to avoid perfect \vspace{-0.2cm} } \\ 
\multicolumn{17}{l}{ \hspace{1.2cm}  multicollinearity } \\ 
\multicolumn{17}{l}{\hspace{1.2cm} $+$ estimate is positive; $-$ estimate is negative  \vspace{-0.22cm}} \\  
\multicolumn{17}{l}{\hspace{1.2cm} $^{\bullet}$p$<$0.1;$^{*}$p$<$0.05; $^{**}$p$<$0.01; $^{***}$p$<$0.001} \\  
\end{tabular}
\end{small}
\end{table} 

In Table~\ref{tCoastals} models estimated using the whole sample are compared with models estimated for the subsample of coastal counties which includes $274$ counties. Columns~$(2) - (5)$ include estimates of models based on the whole sample and they are the same as the estimates in Table~\ref{tPeriods1}. Estimates of models based on subsample of coastal counties are in columns~$(6)$ and $(7)$ in Table~\ref{tCoastals}. These models do not need spatial correction, therefore equation~\eqref{Third} is used. The models for coastal counties do not include distance from coast either.

\begin{table} \centering
\caption{\textbf{Sea level rise and coast distance estimates}}\label{tCoastals}
\begin{center}
\begin{tabular}{|l|cc|cc|cc|cc|cc|cc|}
\hline
& \multicolumn{8}{c|}{\textbf{All counties}} & \multicolumn{4}{c|}{\textbf{Coastal counties}} \\
& \multicolumn{8}{c|}{\textbf{SAR models~\eqref{SAR}}} & \multicolumn{4}{c|}{\textbf{3SLS}} \\
\cline{2-13}
\multicolumn{1}{|c|}{\textbf{Period }} & \multicolumn{4}{c|}{\textbf{SLR }}  & \multicolumn{4}{c|}{\textbf{Coast}} & \multicolumn{4}{c|}{\textbf{SLR}}  \\

\textbf{ } & \multicolumn{4}{c|}{\textbf{ }} &  \multicolumn{4}{c|}{\textbf{distance}} & \multicolumn{4}{c|}{} \\
\cline{2-13}
&\multicolumn{2}{l|}{\textbf{Linear}}& \multicolumn{2}{c|}{\textbf{Sq.}} &\multicolumn{2}{l|}{\textbf{Linear}} & \multicolumn{2}{c|}{\textbf{Sq.}} &\multicolumn{2}{c|}{\textbf{ Linear}} & \multicolumn{2}{c|}{\textbf{Sq.}} \\
\hline 

$1990-2012$& + & * & - & & - & *** & + & ***&+ & $\bullet$& - & \\
$1990-2011$ & + & & +& &- &&+&***& + & &- & \\
$1990-2010$ & + & $\bullet$ & -& & - & ** &+&*** & +& & +&\\
$1990-2009$ & + & *** & - & ** & - & ** & + & *** & +&* &- & \\
$1990-2008$ & - & & + & &- & $ \bullet $ & +& **& +& & +& \\
$1990-2007$ & - & & + & & - & & + & * &+ & &+& \\
$1990-2006$ & + & *** & - & ** & - & $\bullet$ & + & * &- &$\bullet$ &+ & * \\
$1990-2005$ & +& *** & - & *** & - & * & + & *** & +& *& -& $\bullet$\\
$1990-2004$ & + & *** & - & *** & - & * & + & ***& +& *& -& * \\
$1990-2003$ & + & & - & & - & ** & + & *** & + & ** & - & * \\
$1990-2002$ & + & *** & - & *** & - &* & + & ** & + & ** & - & * \\
$1990-2001$ & + & *** & - & *** & - & ** & + & *** & + & ** & - & * \\
$1990-2000$ & + & *** & - & *** & - & ** & + & *** & + & *& -& * \\
\hline
\multicolumn{1}{|l|}{Observations:} & \multicolumn{8}{r|}{$3063$} & \multicolumn{4}{r|}{$274$} \\ 
\hline
\multicolumn{1}{l}{\textit{Notes:}} & \multicolumn{12}{l}{All models include all covariates from~Table~\ref{t3SLSSARall} except of \vspace{-0.2cm}} \\ 
\multicolumn{1}{l}{ }& \multicolumn{12}{l}{ coast distance variables which are not included in the model \vspace{-0.2cm}} \\ 
\multicolumn{1}{l}{ } & \multicolumn{12}{l}{ for the coastal counties and dummy variables for the following \vspace{-0.3cm}} \\ 
\multicolumn{1}{l}{ } & \multicolumn{12}{l}{regions: Great Lakes, Plains, Southwest and Rocky Mountain, \vspace{-0.2cm}} \\ 
\multicolumn{1}{l}{ } & \multicolumn{12}{l}{ which are not included in the models for the coastal counties  \vspace{-0.3cm}} \\ 
\multicolumn{1}{l}{ } & \multicolumn{12}{l}{ to avoid perfect multicollinearity} 
\vspace{0.15cm} \\
\multicolumn{1}{l}{} & \multicolumn{12}{l}{$+$ estimate is positive; $-$ estimate is negative \vspace{-0.2cm}} \\ 
\multicolumn{1}{l}{} & \multicolumn{12}{l}{$^{\bullet}$p$<$0.1;$^{*}$p$<$0.05; $^{**}$p$<$0.01; $^{***}$p$<$0.001} \\ 
\end{tabular}
\end{center}
\end{table} 

We can see in Tables~\ref{tPeriods} and~\ref{tCoastals} that both quadratic and linear sea level rise terms are only highly significant when the models are estimated for all counties. As displayed in Table~\ref{tPeriods}, the sea level rise terms are not significant at all for almost all models of the near coast counties while they remain slightly significant in models for coastal counties in Table~\ref{tCoastals}, which do not include the coast distance terms. This suggests that the reason why the sea level rise coefficients are significant in models for all counties, is because they partially capture the effects of distance from the coast.

\FloatBarrier
\subsection{Government finances}\label{GovFin}
The government finances variables are important as coastal protection is usually funded by federal, state or county government. As we can see in Table~\ref{t3SLSSAR}, the estimates of per capita local tax income and per capita highway and education expenditures have different signs than expected. The estimate of per capita local tax income is positive and highly significant, and the estimate of per capita highway and education expenditures is negative and insignificant. 

Previous research, for example~\citet{TaxBartik} and~\citet{TaxBecsi}, indicates that the state and local tax income have negative and statistically significant effects on economic growth. Reverse causality is one explanation for the opposite sign of tax income. In richer counties more taxes are paid, so it might appear as if higher taxes cause higher economic growth. Another explanation is the existence of one or more omitted covariates which are correlated with per capita local tax income and per capita income growth. The omitted variables can be other government expenditures and taxes not captured in the model. According to~\citet{TaxHelms}, the positive impact on location and production provided by improved quality of services can be higher than negative impact of higher taxes when the revenue from taxes is used to finance public services.~
This can also explain the positive sign of the local tax income coefficient.

Comparing estimates of per capita tax income for the $13$ time periods, it turns out that the positive and significant effect is not consistent over time. As we can see in Table~\ref{tPeriodsFinances}, the coefficient is negative and significant in two cases and in two other cases it is negative and insignificant. 

\hspace*{-4.27cm}\begin{table} \centering
\caption{\textbf{ Estimates of local government finances variables:}}\label{tPeriodsFinances}
\caption*{SAR models~\eqref{SAR} for different time periods}
\begin{center}
\begin{tabular}{|l|cc|cc|}
\hline
\multicolumn{5}{|c|}{\textbf{Local government finances variables (per capita)}} \\
\hline
& \multicolumn{2}{c|}{\textbf{Direct expenditures  }} & \multicolumn{2}{l|}{} \\
\multicolumn{1}{|c|}{\textbf{Period}} &\multicolumn{2}{c|}{\textbf{ for highways }} & \multicolumn{2}{c|}{\textbf{Total taxes}}  \\
&\multicolumn{2}{c|}{\textbf{ and education }} & \multicolumn{2}{l|}{ }  \\
\hline 
$1990-2012$& - &  & + & *** \\
$1990-2011$ & + & & - &  \\
$1990-2010$ & - &  & +& *\\
$1990-2009$ & - & *** & + & *** \\
$1990-2008$ & - & & - & *** \\
$1990-2007$ & - & & - & *** \\
$1990-2006$ & - & *** & + & ***  \\
$1990-2005$ & -& *** & + & *** \\
$1990-2004$ & - & *** & + & ***  \\
$1990-2003$ & - & & - &  \\
$1990-2002$ & - & *** & + & ***  \\
$1990-2001$ & - & *** & + & ***  \\
$1990-2000$ & - & *** & + & ***  \\
\hline
\multicolumn{1}{|l|}{Observations:} & \multicolumn{4}{r|}{$3063$} \\ 
\hline
\multicolumn{1}{l}{\textit{Notes:}} & \multicolumn{4}{r}{All models include all covariates from~Table~\ref{t3SLSSARall}} \\ 
\multicolumn{1}{l}{} & \multicolumn{4}{r}{$+$ estimate is positive; $-$ estimate is negative} \\ 
\multicolumn{1}{l}{} & \multicolumn{4}{r}{$^{\bullet}$p$<$0.1;$^{*}$p$<$0.05; $^{**}$p$<$0.01; $^{***}$p$<$0.001} \\ 
\end{tabular}
\end{center}
\end{table}

The negative sign of per capita highway and education expenditures which was obtained by~fitting~\eqref{SAR} for the longest time period $1990-2012$ also contradicts our expectations. However, as we can see in Table~\ref{tPeriodsFinances}, for almost half of the time periods including the longest one the coefficient is not significant and in one case it is positive. The negative and significant estimates of the other periods could be explained by the existence of one or more omitted covariates which are correlated with per capita government expenditures and per capita income growth similarly as~in~the~case of~per~capita tax income.

Because the government finances and their effects on economic growth are not the main focus of~this study, we decided not to search for all of the data which would reflect the government finances more accurately. Instead, we estimated model~\eqref{SAR} without the government finances variables and we also estimated several variants~of~\eqref{SAR} which include other local government revenue variables instead of per capita tax income to verify whether the results remain robust. The per capita highway and education expenditures variable is omitted in some of these variants. The signs and significance levels of the estimates of sea level rise and local government finances variables of these variants are summarised in Table~\ref{tFinances}. The economic growth rate variable in all models in~Table~\ref{tFinances} reflects time period~$1990-2012$. Each row represents one variant and all government finance variables are per capita, for fiscal year 1992. Though we estimated each variant for all $13$ time periods and each of these models include also all other covariates from Table~\ref{t3SLSSARall} (except of~government expenditures and tax income unless listed in~Table~\ref{tFinances}), estimates of the other periods and the other coefficients are not presented here to save space as they do not differ substantially. The first row represents the same specification as the second column of Table~\ref{t3SLSSAR} and it is included for comparison.

\begin{table} \centering
\begin{footnotesize}
\footnotesize
\caption{\textbf{Sea level rise and coast distance estimates:}}\label{tFinances} 
\vspace{-0.27cm}
\caption*{SAR models~\eqref{SAR} with various local government finances variables}
\begin{tabular}{|ll|cc|cc|cc|cc|}
\hline
\multicolumn{2}{|c|}{\textbf{Government finances variables included}} & \multicolumn{8}{c|}{ \textbf{Period $1990-2012$}} \\
\cline{3-10}
\multicolumn{2}{|c|}{\textbf{(all per capita)}}  & \multicolumn{2}{l|}{\textbf{SLR}} & \multicolumn{2}{l|}{\textbf{SLR}} & \multicolumn{4}{c|}{\textbf{Government finances}}  \\
\cline{1-2}
\cline{7-10}
\textbf{Direct Expenditures}&\textbf{General Revenue}& & & \multicolumn{2}{c|}{\textbf{sq.}} &\multicolumn{2}{c|}{\textbf{Exp.}} & \multicolumn{2}{c|}{\textbf{Revenue}} \\
\hline 
For highway and education & Total taxes & +&* &- & &- & &+& *** \\
For highway and education & Total intergov. & + & ** & - & & + & *** &- &***\\
\multirow{2}{*}{For highway and education} & Intergovernmental & \multirow{2}{*}{+} & \multirow{2}{*}{**} & \multirow{2}{*}{-} & & \multirow{2}{*}{+} & \multirow{2}{*}{***} &\multirow{2}{*}{-} &\multirow{2}{*}{***}\\
\vspace{-0.9cm} \\
& \hspace{0.3cm} from state gov. &  &  &  & &  &  & &\\
\multicolumn{1}{|c}{\tiny{$---$}} & Total taxes & + & * & - & & \tiny{$---$} & \tiny{$---$} & + & ***\\
\multicolumn{1}{|c}{\tiny{$---$}}& Total intergov. & +& ** & - & & \tiny{$---$} & \tiny{$---$} & - & *** \\
\multicolumn{1}{|c}{\multirow{2}{*}{\tiny{$---$}}}& Intergovernmental & \multirow{2}{*}{+}& \multirow{2}{*}{**} & \multirow{2}{*}{-} & & \multirow{2}{*}{\tiny{$---$}} & \multirow{2}{*}{\tiny{$---$}} & \multirow{2}{*}{-} & \multirow{2}{*}{***}\\
\vspace{-0.9cm} \\
& \hspace{0.3cm} from state gov. & & &  & &  &  &  & \\
\multicolumn{1}{|c}{\tiny{$---$}} & Property taxes &+& * & - & & \tiny{$---$} & \tiny{$---$} & + & ***\\
\multicolumn{1}{|c}{\tiny{$---$}}& \multicolumn{1}{c|}{\tiny{$---$}} &+& ** & - & & \tiny{$---$} & \tiny{$---$} & \tiny{$---$} & \tiny{$---$} \\
\hline
\multicolumn{2}{|c|}{Observations:} & \multicolumn{8}{r|}{$3063$}  \\ 
\hline

  \multicolumn{10}{p{15.5cm}}{ \textit{Notes:} \hspace{0.6cm} All models include all covariates from~Table~\ref{t3SLSSARall} (except of government expenditures }  \\ 
    \multicolumn{10}{p{15.5cm}}{\vspace{-0.7cm} \hspace{1.8cm} and tax income unless listed in the table}  \\ 
 \multicolumn{10}{p{15.5cm}}{ \hspace{1.8cm} \begin{tiny} $---$ \end{tiny} if no government finances variable included;~ $^{\bullet}$p$<$0.1;$^{*}$p$<$0.05; $^{**}$p$<$0.01; $^{***}$p$<$0.001} \\ 
\multicolumn{10}{p{15cm}}{ \hspace{1.8cm} $+$ estimate is positive; $-$ estimate is negative} \\ 
\end{tabular}
\end{footnotesize}
\end{table}

 Sea level rise and coast distance coefficients obtained by fitting two variants of spatial autoregressive model~\eqref{SAR} are summarized and compared in~Table~\ref{tNoFinance}. The variant in columns~$(2)~-~(5)$ was obtained by fitting our basic variant of~\eqref{SAR} with all covariates including total per capita taxes and per capita highway and education expenditures  and the one in columns~$(6)~-~(9)$ was obtained by~\eqref{SAR} with all covariates excluding the government finances variables. We can see that the signs and significance levels do not differ for most periods.

\begin{table} \centering
\begin{small}
\caption{\textbf{SAR models~\eqref{SAR}: Sea level rise and coast distance estimates}}\label{tNoFinance} 
\caption*{{Comparison of models with and without local government finances variables}}
\begin{tabular}{|c|cc|cc|cc|cc|cc|cc|cc|cc|}
\hline
& \multicolumn{8}{c|}{\textbf{Including per capita taxes}} & \multicolumn{8}{c|}{\textbf{ Without per capita taxes}} \\
& \multicolumn{8}{c|}{\textbf{and expenditures for highways}} & \multicolumn{8}{c|}{\textbf{and expenditures for highways}} \\
& \multicolumn{8}{c|}{\textbf{and education}} & \multicolumn{8}{c|}{\textbf{and education}} \\
\cline{2-17}
\textbf{Period } & \multicolumn{4}{c|}{\textbf{SLR }}  & \multicolumn{4}{c|}{\textbf{Coast}} & \multicolumn{4}{c|}{\textbf{SLR}}  & \multicolumn{4}{c|}{\textbf{Coast}} \\

\textbf{ } & \multicolumn{4}{c|}{\textbf{ }} &  \multicolumn{4}{c|}{\textbf{distance}} & \multicolumn{4}{c|}{} &  \multicolumn{4}{c|}{\textbf{distance}}\\
\cline{2-17}
&\multicolumn{2}{l|}{\textbf{Linear}}& \multicolumn{2}{c|}{\textbf{Sq.}} &\multicolumn{2}{l|}{\textbf{Linear}} & \multicolumn{2}{c|}{\textbf{Sq.}} &\multicolumn{2}{c|}{\textbf{ Linear}} & \multicolumn{2}{c|}{\textbf{Sq. }} & \multicolumn{2}{c|}{\textbf{ Linear}} & \multicolumn{2}{c|}{\textbf{Sq.}} \\
\hline 
$1990-2012$& + & * & - & & - & *** & + & *** & + & ** & - & & - & *** & +& ***\\
$1990-2011$ & + & & +& &- &&+&*** & - & & + & & - & & + & ***\\
$1990-2010$ & + & $\bullet$ & -& & - & ** &+&*** & + & $\bullet$ & -& &- & ** &+&***\\
$1990-2009$ & + & *** & - & ** & - & ** & + & *** & + & *** & - & ** & - & ** & + & ***  \\
$1990-2008$ & - & & + & &- & $ \bullet $ & +& ** & - & & + & &- &  & +& ** \\
$1990-2007$ & - & & + & & - & & + & * & - & & + & & - & & + & $\bullet$  \\
$1990-2006$ & + & *** & - & ** & - & $\bullet$ & + & * & + & *** & - & ** & - & $\bullet$ & + & *  \\
$1990-2005$ & +& *** & - & *** & - & * & + & *** & +& *** & - & *** & - & * & + & *** \\
$1990-2004$ & + & *** & - & *** & - & * & + & ***  & + & *** & - & *** & - & * & + & ** \\
$1990-2003$ & + & & - & & - & ** & + & ***& + & & - & & - & ** & + & ***  \\
$1990-2002$ & + & *** & - & *** & - &* & + & ** & + & *** & - & *** & - &* & + & * \\
$1990-2001$ & + & *** & - & *** & - & ** & + & *** & + & *** & - & *** & - & ** & + & ** \\
$1990-2000$ & + & *** & - & *** & - & ** & + & *** & + & *** & - & *** & - & ** & + & **\\
\hline
\multicolumn{1}{|c|}{Obs.:} & \multicolumn{8}{r|}{$3063$} & \multicolumn{8}{r|}{ $3063$} \\ 
\hline
\multicolumn{1}{l}{\textit{Notes:}} & \multicolumn{16}{p{13cm}}{All models include all covariates from~Table~\ref{t3SLSSARall} (except of the government finances variables for the second model)} \\ 
\multicolumn{1}{l}{} & \multicolumn{16}{p{13cm}}{$+$ estimate is positive; $-$ estimate is negative;~ $^{\bullet}$p$<$0.1;$^{*}$p$<$0.05; $^{**}$p$<$0.01; $^{***}$p$<$0.001} \\  
\end{tabular}
\end{small}
\end{table} 

Estimates of all coefficients of~the~spatial autoregressive model~\eqref{SAR} without any government finances variables are summarized in~Table~\ref{tNoFinCompare} in~Appendix~3. The period of economic growth of this model is \mbox{$1990-2012$}. We can see that the estimates are similar to our basic variant in the second column of~Table~\ref{t3SLSSAR}. Also the coefficients of the other specifications from~Table~\ref{tFinances} are very similar as well as its estimates for the other time periods. However, these are not presented in this paper to keep its length within reasonable limit. 

We can conclude that the estimates are reasonably robust with respect to government finances variables.

\FloatBarrier
\section{Conclusion}\label{sConclusion}
A common assumption in numerous studies is that sea level rise has negative effects on the economy. Here, in the first empirical test, we did not find a statistically robust and significant effect of sea level rise on economic growth in the continguous USA \textemdash if anything, the estimated impact is positive.

A growth model and a matching estimator were used to investigate the effects of sea level rise on the economy of the United States. We applied a 3SLS method with spatial correction to estimate the economic growth model. The model was estimated for~$13$ different time periods, each of them starting in year 1990 and ending in a year between 2000 and 2012. In some of these models, in particular for period 1990-2006 and some shorter periods, we found a statistically significant relationship, however it is not present for all periods. In almost half of the models presented in Table~\ref{tPeriods1} both sea level rise coefficients are insignificant. Further, different variants of the economic growth model were estimated to verify whether the results remain unchanged. We found that in models for near coast and coastal counties the sea level rise coefficients are less significant and they are not significant at all in models without outliers. Hence, the occasional significant effects may be driven by outliers, or may be statistical flukes. The results of the other robustness tests do not differ substantially from the estimates of spatial autoregressive models~\eqref{SAR} presented in Tables~\ref{tPeriods1} and~\ref{t3SLSSARall}. We used three different matchings that are balanced on all relevant covariates in our dataset. The estimated treatment effect is insignificant in all three cases, which is in accordance with the results of the economic growth model. There is therefore no statistically discernible impact of past sea level rise on economic growth in the USA.

One reason why we did not find a stable significant effect may be the fact that sea level rise is a~gradual and slow process, developing over decades and centuries if not millennia, and its effects can be apparent only for a longer time period. The longest period for which the effects are analysed in this study is $22$ years. A logical continuation of this study would be an extension long-term growth, however data from more than $60$ or $70$ years ago are hardly available for all required covariates. A possible solution could be the use of sparse regression without the unavailable covariates. This is a topic for future research.

Instead of economic growth, alternative indicators could be used, such as land prices as it is plausible that they are affected by sea level rise, or the composition of public investment as that is plausibly affected by coastal protection.

It may also be that, as with other impacts of climate change, sea level rise has a minimal effect on a developed economy like that of the USA, but a more substantial impact on less developed economies. In order to test this hypothesis, the current study would need to be repeated either for currently poor countries or for sea level rise in the distant past. In either case, data availability may be a real problem.

However, as it stands, no stable, significant effect of sea level rise on economic growth was found. More research should be done on this topic as possible significant effects could be found for different regions or different time periods, but for now that is the conclusion.

\FloatBarrier

\bibliographystyle{apa} 
\makeatletter 
\renewcommand{\thesection}{\hspace*{-1.0em}}
\pagebreak

\bibliography{literature}

\pagebreak
\section{Appendix 1 Control variables}

\setcounter{table}{0} 
\makeatletter 
\renewcommand{\thetable}{A\@Roman \c@table} 

\linespread{1.54}
The covariates used in this study are listed in Table~\ref{tVariables}.
\begin{table} \centering
\caption{\textbf{List of Covariates and their description}}\label{tVariables}
\vspace{-0.5cm}
\begin{tabular}{ll}
\multicolumn{2}{l}{\textbf{Government finances}}\\
   \vspace{-0.9cm} \\
Gov. expenditures p. capita & Per capita highway and education expenditures 1992 \\
Tax income per capita& Per capita local tax income 1992  \\
\multicolumn{2}{l}{\textbf{Measures of agglomeration}}\\
   \vspace{-0.9cm} \\
Population density & Population density 1990 \\ 
Urban & Metropolitan counties \\
Rural & Rural counties not adjacent to metropolitan areas \\
\multicolumn{2}{l}{\textbf{Measures of religious adherence}}\\
   \vspace{-0.9cm} \\
Adherents & Per capita total number of religious adherents 1990 \\
Catholics & Per capita Catholics adherents 1990 \\
Evangelical Protestants & Per capita Evangelical Protestants adherents 1990 \\
Mainline Protestants & Per capita Mainline Protestants adherents 1990 \\
Religious diversity & Religious diversity index 1990 \\
\multicolumn{2}{l}{\textbf{Other socioeconomic and environmental
indicators}}\\
   \vspace{-0.9cm} \\
Coast distance & Distance from coast \\
\multirow{2}{*}{Education} & Percent of population (25 years or older) \\
& \hspace{0.5cm} who have bachelor's degree or higher 1990 \\
Highway & Presence of interstate highway interchange \\
Right to work laws & Right to work laws \\
Nonwhites & Percent of population who are nonwhite 1990 \\ 
\multirow{2}{*}{Amenities} & Natural amenities index by~\citet{Amenities} \\
& \hspace{0.3cm} (viz note below table) \\
\multicolumn{2}{l}{\textbf{Regional dummy variables}}\\
   \vspace{-0.9cm} \\
New England & New England region \\
Mideast & Mideast region \\
Great Lakes & Great Lakes region \\
Plains & Plains \\
Southeast & Southeast region \\
Southwest & Southwest region \\
Rocky Mountain & Rocky Mountain region \\
\hline

\multicolumn{2}{p{15cm}}{ \linespread{1} \textit{Note:}  \vspace{-0.3cm} Environmental qualities captured by the natural amenities index: January temperature, Days of sun in January, July temperature, \vspace{-0.3cm} July humidity, Proportion of~water area, Topography}
\end{tabular}
\end{table} 

\linespread{1.6}

Population density and urban and rural dummy variables are included as measures of agglomeration as it is assumed that economic activities are attracted to metropolitan areas which further enhance economic growth.

\citet{Rupasingha} show that the percentage of religious adherents has a significant impact on economic growth as well as the percentages of adherents of individual religious denominations and religious diversity. Similarly, as in~\citet{Rupasingha}, we first considered two specifications, specifically a model with percentage of all religious adherents and a model without this variable, which includes percentages of adherents of the three main denominations, namely Catholics, Evangelical Protestants and Mainline Protestants. The religious diversity index is included in both these specifications. Finally, we chose the second specification as for the first specification both parameters~$\rho$ and~$\lambda$ are significant in the form~\eqref{genSpat} and also according to the LM diagnostic tests for spatial dependence \citep{Bera96} the form~\eqref{genSpat} is correct, but the Moran's I adjusted for residuals is significant for this specification. On the other hand, appropriate specification of the model with the percentages of the three main religious adherents is~\eqref{SAR} ($\lambda$ is insignificant in form~\eqref{genSpat}) and the Moran's I statistic applied to residuals from this model is insignificant.

 The three denominations, specifically Catholics, Evangelical Protestants and Mainline Protestants include most of the $133$ Judeo-Christian church bodies listed in the Yearbook of American and Canadian Churches which responded to the invitation to participate in the study organized by~the~Association of~Statisticians of~American Religious Bodies (ASARB) in 1990. The excluded group includes all other church groups and non-affiliates. Percentage of religious adherents, percentage of Evangelical Protestant adherents and percentage of Mainline protestant adherents are all negatively correlated with dummy variable interstate highway access. Their Pearson's product - moment correlation coefficients are $-0.103 $, $-0.124 $ and $-0.074 $, respectively with both-sided \textit{p}-values $1.009 \times 10^{-8} $, $5.25 \times 10^{-12} $ and $4.523 \times 10^{-5} $, respectively. On the other hand, the percentage of Catholic adherents is weakly positively correlated with highway access dummy variable. Its value of the Pearson's product - moment correlation coefficient is $0.045 $ and the \textit{p}-value is $0.014 $. Since highway construction is usually funded from the same sources as the construction of flood dikes, it is plausible that the percentage of Catholics is positively correlated with construction of dikes, while the percentage of Protestants is negatively correlated with construction of dikes. Therefore the religious variables are relevant and they are included in the model. Religious diversity is included as according to some studies, for example~\citet{Reldiv}, higher religious diversity is related to higher quality religion due to higher competition. On the other hand, in the presence of greater religious plurality societies have less social capital which may lead to a less trusting society and slower economic growth. 
The religious diversity index was obtained similarly as in~\citet{Rupasingha} according to formula
\begin{equation}\label{reldiv}
Reldiv = 1 - \sum_{i=1}^{133}(Denom_i^2),
\end{equation}
where $Denom_i$ denotes share of adherents of denomination $i$.

Education is measured as the percentage of the population who are 25 years or older and have a bachelor's degree or higher. This variable serves as a proxy for human capital. Interstate highway access is a dummy variable which is equal to~$1$ for counties which have interstate highway interchange and~$0$ for other counties and it is included to capture accessibility of counties. Effects of right to work law on the economy and its growth have been studied extensively. In the absence of right to work laws, legislation favours labour unions which raises labour costs and discourages employers from investing. According to some studies, for example~\citet{GrowthRTW} or~\citet{RTW}, there is evidence that right to work laws have a positive and significant effect on economic growth, therefore a state level dummy variable which indicates the presence of right to work laws is included. Percentage of
nonwhite population was found to be associated with earning rates and overall costs of production
by many labour studies therefore it is also included.

 It is further expected that a higher level of natural amenities is related to higher economic growth, thus the natural amenities index derived by~\citet{Amenities} is included. The index is constructed using six measures of climate, topography and water area which are explained in detail in~\citet{Amenities}.

The last seven covariates in Table~\ref{tVariables} are regional dummy variables included to capture regional effects. The omitted region is Far West. 
\FloatBarrier
 \pagebreak

\section{Appendix 2 Data}

Descriptive statistics of sea level rise, average growth rate of per capita income, coast distance, per capita government expenditures and per capita tax income can be found in Table~\ref{tDescriptive} in Section~\ref{Data}. Descriptive statistics of the other covariates are summarized in Table~\ref{tDescriptiveA} below. 

Per capita highway and education expenditures, per capita local tax income, population density, education and percent of population who are nonwhite were obtained from the~\citet{Census}. Urban and rural dummy variables were constructed in the same way as in~\citet{Rupasingha} based on Rural-Urban Continuum Codes, which are published by~\citet{USDA}~(USDA). Variable urban is equal to~$1$ for metropolitan counties with Rural-Urban Continuum Codes~$0-3$ and variable rural is equal to~$1$ for counties with Rural-Urban Continuum Codes~$5$,~$7$ and~$9$ that are not adjacent to metropolitan areas. The excluded group includes rural counties adjacent to metropolitan areas with Rural-Urban Continuum Codes~$4$,~$6$ and~$8$.

The religious variables are available online by the~\citet{ARDA} (ARDA). The data set provided by ARDA contains percentages of religious adherents of $133$ religious denominations who responded to an invitation to participate in the study organized by ASARB in year 1990. The invitation was sent to $246$ denominations that included all Judeo-Christian church bodies listed in the Yearbook of American and Canadian Churches, plus a few others for whom addresses could be found. The $133$ denominations were grouped into three groups, in particular Catholics, Evangelical Protestants and Mainline Protestants in the same way as~\citet{Rupasingha}. These three groups include almost all $133$ participating denominations, the rest is in the excluded category.

\newpage

\begin{table}[H] \centering
\caption{\textbf{Descriptive Statistics }}\label{tDescriptiveA}
\begin{tabular}{lrr}
\hlinewd{1pt}
\textbf{Variable}  & \textbf{Mean} & \textbf{Std. dev.} \\ 
\hlinewd{1pt}
Population density (Rate per square mile) & $166.5973 $& $877.9581 $\\ 
Urban $(0,1)$ & $ 0.2635 $ & $ 0.4406 $ \\
Rural $(0,1)$ & $ 0.4146 $ & $0.4927 $ \\
\hline
\multicolumn{3}{c}{\textbf{Measures of religious adherence}}\\
\hline
Adherents  (Percentage) & $ 59.7319$ & $ 19.8822$ \\
Catholics  (Percentage) & $ 13.0005$ & $ 15.1542$ \\
Evangelical Protestants (Percentage)& $31.4110 $ &$20.5496$\\
Mainline Protestants (Percentage) & $12.9707$& $8.6508$ \\
Religious diversity (Formula~\eqref{reldiv})\vspace{-0.3cm} & \multirow{2}{*}{$ 0.8697$} & \multirow{2}{*}{$ 0.1296 $} \\
\hspace{0.3cm}  \citet{Rupasingha} & &  \\
\hline
\multicolumn{3}{c}{\textbf{Other socioeconomic and environmental
indicators}}\\
\hline
Education (Percentage) & $ 13.3918 $& $ 6.4250 $ \\
Highway  $(0,1)$ & $0.4084 $ & $0.4916 $ \\
Right to work laws $(0,1)$ & $0.6202 $ & $0.4853 $ \\
Nonwhites (Percentage) & $12.7202$ & $15.4563$ \\ 
Amenities (Scale \citet{Amenities}) & $ 0.0505 $ & $ 2.2876 $ \\
\hline
\multicolumn{3}{c}{\textbf{Regional dummy variables}}\\
\hline
New England  $(0,1)$ & $0.0219$ & $0.1463$ \\
Mideast  $(0,1)$ & $0.0568$ & $0.2315$ \\
Great Lakes  $(0,1)$ & $0.1423$ & $0.3495$ \\
Plains  $(0,1)$ & $0.2018$ & $0.2018$ \\
Southeast  $(0,1)$ & $0.3356$ & $0.4723$ \\
Southwest  $(0,1)$ & $0.1224$ & $0.3278$ \\
Rocky Mountain  $(0,1)$ & $0.0702$ & $0.2555$ \\
\hline
\end{tabular}
\end{table}
\FloatBarrier
 \pagebreak

\section{Appendix 3 Tables}

\vspace{1.5cm}
\begin{table}[H] 
\caption{}\label{tGrowthDes}
\begin{center}
\begin{tabular}{p{1.25cm}p{1.25cm}p{3.5cm}p{3cm}p{4cm}}
\multicolumn{5}{c}{\textbf{Average growth rate of per capita income $\boldsymbol{g_n}$, various time periods:}} \\ 
\multicolumn{5}{c}{\underline{\textbf{Descriptive statistics}}} \\ 
\\
 & &  \multicolumn{1}{l}{\hspace{0.3cm} \underline{\textbf{Period}}} & \multicolumn{1}{l}{\hspace{-0.05cm}\underline{\textbf{Mean}}} & \multicolumn{1}{l}{\hspace{-1.35cm} \underline{\textbf{Standard deviation}}} \\ 
& & $1990-2012$ & $0.0413$ & $0.0076$ \\ 
& & $1990-2011$ & $0.0415$  & $0.0075$ \\ 
& & $1990-2010$&$0.0402$ & $0.0070$ \\ 
& & $1990-2009$&$0.0408$ & $0.0072$ \\ 
& & $1990-2008$& $0.0443$ & $0.0075$ \\ 
& & $1990-2007$& $0.0435$ & $0.0069$ \\ 
& & $1990-2006$& $0.0423$ & $0.0074$ \\ 
& & $1990-2005$&$0.0427$ & $0.0071$ \\ 
& & $1990-2004$&$0.0429$ & $0.0076$ \\ 
& & $1990-2003$&$0.0425$ & $0.0077$ \\ 
& & $1990-2002$&$0.0418$ & $0.0085$ \\ 
& & $1990-2001$&$0.0453$ & $0.0088$ \\ 
& & $1990-2000$& $0.0439$ & $0.0098$ \\ 
\end{tabular}
\end{center}
\end{table}

\newpage
\begin{table}[H] \centering
\caption {\small \hspace{0.3cm} \textbf{OLS~\eqref{growth}, \hspace{0.15cm}} \textit{Growth rate between 1990-2012}
 \hfill \null~} 
\vspace{-0.7cm}
\label{tOLS} 
\small 
\begin{tabular}{@{\extracolsep{-1pt}}lD{.}{.}{-4} } 
\vspace{-0.5cm} \\
\hline \\
\vspace{-1.5cm} \\
Constant & 0.2250$ $(0.0084)^{***} \\ 
Log of initial income pp. (US\$)  & -0.0200$ $(0.0009)^{***} \\ 
Sea level rise (m/year) & 0.5337$ $(0.2677)^{*} \\ 
Sea level rise (m/year) - squared & -18.3600$ $(35.7000) \\ 
Coast distance (thousands km) & -0.007$ $(0.001)^{***} \\ 
Coast distance (thousands km) - squared & 0.008$ $(0.001)^{***} \\ 
Gov. expenditures per capita ( US\$) & -0.3145$ $(0.4336) \\
Tax income per capita (bn. US\$) & 2.4300$ $(0.4001)^{***} \\ 

  \multicolumn{2}{c}{\textbf{Measures of agglomeration}}\\

Population density (rate per thousand square miles)& 0.0920$ $(0.1370) \\ 
Urban (dummy) & 0.00002$ $(0.0003) \\ 
Rural (dummy) & 0.0005$ $(0.0003) \\ 

  \multicolumn{2}{c}{\textbf{Measures of religious adherence}}\\

Catholics (percentage)& 0.0001$ $(0.00001)^{***} \\ 
Evangelical Protestants (percentage)& 0.0001$ $(0.00001)^{***} \\ 
Mainline Protestants (percentage)& 0.0001$ $(0.00002)^{**} \\ 
Religious diversity (Formula~\eqref{reldiv})& 0.0031$ $(0.0012)^{*} \\ 

  \multicolumn{2}{c}{\textbf{Other socioeconomic and environmental indicators}}\\

Education (percentage) & 0.0002$ $(0.00003)^{***} \\ 
Highway (dummy) & -0.0004$ $(0.0002) \\ 
Right to work laws (state level dummy) & 0.0012$ $(0.0003)^{***} \\ 
Nonwhites (percentage) & -0.00004$ $(0.00001)^{***} \\ 
Amenities (scale \citet{Amenities}) & -0.0003$ $(0.0001)^{***} \\ 
  \multicolumn{2}{c}{\textbf{Regional dummy variables}}\\
New England (dummy)& -0.0006$ $(0.0010) \\ 
Mideast (dummy)& -0.0017$ $(0.0008)^{*} \\ 
Great Lakes (dummy)& -0.0045$ $(0.0009)^{***} \\ 
Plains (dummy)& -0.0027$ $(0.0009)^{**} \\ 
Southeast (dummy)& -0.0033$ $(0.0007)^{***} \\ 
Southwest (dummy)& 0.0001$ $(0.0008) \\ 
Rocky Mountain (dummy)& -0.0012$ $(0.0008) \\ 
\vspace{-0.6cm}\\
\hline
\multicolumn{2}{l}{\vspace{-0.5cm} \textit{Notes:} \hspace{0.5cm}$^{*}$p$<$0.05; $^{**}$p$<$0.01; $^{***}$p$<$0.001, \hspace{0.5cm}Standard errors in brackets} \\ 
\end{tabular} 
\end{table} 

\newpage

Adjusted R-squared is~$0.374$ for the OLS estimate of regression~\eqref{growth} in Table~\ref{tOLS} and the \textit{F}-statistic is $71.36$ which is significant with a \textit{p}-value lower that $2.2 \times 10^{-16}$.
\\

\vfill

\begin{table}[H] \centering
\caption{ \textbf{3SLS - first and second stage}, \hfill \textit{Growth rate between 1990-2012} } 
\label{tStage12} 
\small 
\begin{tabular}{@{\extracolsep{-1pt}}lcc} 
\hline 
\hline \\[-1.8ex] 
& {Stage 1 eq.~\eqref{FirstSt}} & {Stage 2 eq.~\eqref{FirstSec}} \\ 
\vspace{0.25cm}  
\textbf{Dependent variable:} & $ {\Delta y_{n,0}}$ & ${\Delta g_n}$ \\
\hline
Constant & 0.0207 (0.0026)$^{***}$ & 0.0010 (0.0003)$^{***}$ \\ 
Religious adherents (percentage) & 0.0006 (0.00005)$^{***}$ & \\ 
Population density (rate per thousand sq. miles) & 0.2887 (0.9410) & \\ 
Predicted log of initial per capita income (US\$)& & $-$0.0333 (0.0049)$^{***}$ \\ 
\hline 
\multicolumn{3}{l}{ \textit{Notes:}\hspace{0.5cm} $^{*}$p$<$0.05; $^{**}$p$<$0.01; $^{***}$p$<$0.001, \hspace{0.5cm} Standard errors in brackets} \\ 
\end{tabular} 
\end{table} 

\vspace{1.2cm}

The~\textit{F}-statistic of the first stage regression in the first column of Table~\ref{tStage12} is $85.82$ and its \textit{p}-value is lower than $2.2 \times 10^{-16}$. The~\textit{F}-statistic of the second stage in the second column of Table~\ref{tStage12} is $46.14$ and the corresponding \textit{p}-value is $1.319 \times 10^{-11}$. Value of Sargan test statistic of over-identifying restrictions in the IV estimation is $0.796$ and its~\textit{p}-value is $0.372$, thus the test is insignificant and the over-identifying restrictions are valid.

\newpage
\linespread{1.54}
\begin{table}[H] \centering 
\small 
\caption{\small \textit{\textbf{\hspace{0.2cm} Spatial autoregressive model~\eqref{SAR}},\hfill Growth rate between 1990-2012 } } 
\label{t3SLSSARall} 
\small 
\hspace{-1.1cm}
\vspace{-0.8cm} \\
\begin{tabular}{@{\extracolsep{-1pt}}lcc} 
\\ [-1.8ex]\hline \\
\vspace{-1.5cm} \\

& \bf{3SLS~\eqref{Third}} & \bf{SAR~\eqref{SAR}} \\ 
Constant & 0.3476 (0.0017)$^{***}$ & 0.1849 (0.0074)$^{***}$ \\
Log of initial per capita income (US\$) & $-$0.0333 (0.0049)$^{***}$ & $-$0.0333 (0.0049)$^{***}$ \\ 
Sea level rise (m/year) &  0.9467 (0.2768) $^{***}$ & 0.5943 (0.2524)$^{*}$ \\ 
Sea level rise (m/year) - squared & $-$59.2000 (37.0400) & $-$44.4060 (33.7110) \\ 
Coast distance (thousands km) & $-$0.0072 (0.0013)$^{***}$ & $-$0.0045 (0.0012)$^{***}$ \\ 
Coast distance (thousands km) - squared & 0.0083 (0.0007)$^{***}$ & 4535.1000 (690.0000)$^{***}$ \\ 
Gov. expenditures per capita (billion US\$) & $-$0.7102 (0.4515) & $-$0.5957 (0.4106) \\ 
Tax income per capita (billion US\$) & 4.1710 (0.3993)$^{***}$ & 3.3698 (0.3681)$^{***}$ \\ 
   $\rho$ (SAR) & --- & 0.4583 (0.0206)$^{***}$ \\   
   \vspace{-1cm} \\
\multicolumn{3}{c}{\textbf{Measures of agglomeration}}\\
Population density (per thousand square miles) & 0.2527 (0.1429) & $-$0.0213 (0.1303)  \\
Urban (dummy) & 0.0012 (0.0003)$^{***}$ & 0.0009 (0.0003)$^{**}$ \\ 
Rural (dummy) & 0.00004 (0.0003) & 0.0003 (0.0003) \\ 
  \multicolumn{3}{c}{\textbf{Measures of religious adherence}}\\
Catholics (percentage) & 0.0001 (0.00001)$^{***}$ & 0.0001 (0.00001)$^{***}$ \\ 
Evangelical Protestants (percentage) & 0.0001 (0.00001)$^{***}$ & 0.0001 (0.00001)$^{***}$ \\ 
Mainline Protestants (percentage) & 0.0001 (0.00002)$^{***}$ & 0.0001 (0.00001)$^{***}$ \\ 
Religious diversity (Formula~\eqref{reldiv}) & 0.0057 (0.0013)$^{***}$ & 0.0039 (0.0012)$^{***}$ \\ 
  \multicolumn{3}{c}{\textbf{Other socioeconomic and environmental indicators}}\\
Education (percentage) & 0.0004 (0.00002)$^{***}$ & 0.0003 (0.00002)$^{***}$ \\ 
Highway (dummy) & $-$0.0002 (0.0003) & $-$0.0001 (0.0002) \\ 
Right to work laws (state level dummy) & 0.0018 (0.0003)$^{***}$ & 0.0010 (0.0003)$^{***}$ \\ 
 Nonwhites (percentage) & $-$0.0001 (0.00001)$^{***}$ & $-$0.0001 (0.00001)$^{***}$ \\ 
Amenities (scale \citet{Amenities})& $-$0.0003 (0.0001)$^{***}$ & $-$0.0002 (0.0001)$^{*}$ \\ 
   \vspace{-0.9cm} \\
  \multicolumn{3}{c}{\textbf{Regional dummy variables}}\\
New England (dummy)& $-$0.0018 (0.0010) & $-$0.0025 (0.0010)$^{**}$ \\ 
Mideast (dummy)& $-$0.0030 (0.0008)$^{***}$ & $-$0.0023 (0.0008)$^{**}$ \\ 
Great Lakes (dummy)& $-$0.0063 (0.0009)$^{***}$ & $-$0.0031 (0.0008)$^{***}$ \\ 
Plains (dummy)& $-$0.0054 (0.0010)$^{***}$ & $-$0.0028 (0.0009)$^{**}$ \\ 
Southeast (dummy)& $-$0.0061 (0.0008)$^{***}$ & $-$0.0026 (0.0007)$^{***}$ \\ 
Southwest (dummy)& $-$0.0031 (0.0008)$^{***}$ & $-$0.0017 (0.0007)$^{*}$ \\ 
Rocky Mountain (dummy)& $-$0.0032 (0.0008)$^{***}$ & $-$0.0020 (0.0008)$^{**}$ \\ 
\hline 
   \vspace{-0.8cm} \\
\multicolumn{3}{l}{\vspace{-0.5cm} \textit{Notes:} \hfill $^{*}$p$<$0.05; $^{**}$p$<$0.01; $^{***}$p$<$0.001, \hspace{0.5cm}Standard errors in brackets} \\ 
\end{tabular} 
\end{table} 

\linespread{1.6}
\newpage

\begin{table}[H] \centering 
\caption{\small \textbf{\hspace{0.2cm} Spatial autoregressive model~\eqref{SAR}} - Impact measures, \hspace{1cm} \textit{1990-2012 }} 
\label{tSARImpacts2} 
\small 
\begin{tabular}{@{\extracolsep{-1pt}}lrrr} 

\vspace{-1.5cm} \\
\\[-1.8ex]\hline 
\vspace{-1.2cm} \\
\\[-1.8ex] &  \textbf{Direct} &  \textbf{Indirect} &  \textbf{Total} \\ 
\vspace{-0.8cm} \\
Sea level rise (m/year) &  0.6218  & 0.4753 & 1.0971 \\ 
Sea level rise (m/year) - squared & $-$46.4611 & $-$35.5122 &  $-$81.9733\\ 
Coast distance (thousands km) & $-$0.0048 & $-$0.0036 & $-$0.0084 \\ 
Coast distance (thousands km) - squared & 4,744.9020 & 3,626.7320 & 8,371.6340 \\ 
Gov. expenditures per capita (billion US\$) & $-$0.6232 & $-$0.4764 & $-$1.0996 \\ 
Tax income per capita (billion US\$) & 3.5257 & 2.6948 & 6.2205\\  
\multicolumn{4}{c}{\textbf{Measures of agglomeration}}\\
Population density (rate per thousand square miles) & $-$0.0223 & $-$0.0171  & $-$0.0394 \\
Urban (dummy) & 0.0009 & 0.0007 & 0.0016 \\ 
Rural (dummy)& 0.0003 & 0.0003 & 0.0006 \\ 
  \multicolumn{4}{c}{\textbf{Measures of religious adherence}}\\
Catholics (percentage) & 0.0001  & 0.0001 & 0.0001 \\ 
Evangelical Protestants (percentage) & 0.0001  & 0.0001 & 0.0001 \\ 
Mainline Protestants (percentage) & 0.0001  & 0.0001 & 0.0001 \\ 
Religious diversity (Formula~\eqref{reldiv}) & 0.0041 & 0.0031 & 0.0072 \\ 
  \multicolumn{4}{c}{\textbf{Other socioeconomic and environmental indicators}}\\
Education (percentage) & 0.0003 & 0.0003 & 0.0006\\ 
Highway (dummy) & $-$0.0001 & $-$0.0001 & $-$0.0002 \\ 
Right to work laws (state level dummy) & 0.0011 & 0.0008 & 0.0019 \\ 
 Nonwhites (percentage) & $-$0.0001 & $-$0.0001  & $-$0.0001\\ 
Amenities (scale \citet{Amenities})& $-$0.0002 & $-$0.0001 & $-$0.0003\\ 
  \multicolumn{4}{c}{\textbf{Regional dummy variables}}\\
New England (dummy)& $-$0.0026 & $-$0.0020 & $-$0.0047 \\ 
Mideast (dummy)& $-$0.0024 & $-$0.0018 & $-$0.0042 \\ 
Great Lakes (dummy)& $-$0.0032 & $-$0.0025 & $-$0.0057 \\ 
Plains (dummy)& $-$0.0030 & $-$0.0023 & $-$0.0052 \\ 
Southeast (dummy)& $-$0.0028 & $-$0.0021 & $-$0.0049  \\ 
Southwest (dummy)& $-$0.0018 & $-$0.0014 & $-$0.0032 \\ 
Rocky Mountain (dummy)& $-$0.0021 & $-$0.0016  & $-$0.0037 \\ 

\hline 
\hline \\[-1.8ex] 
\end{tabular} 
\end{table}

\newpage
\linespread{1.55}
\begin{table}[H] \centering 
  \caption{~~~ \textit{Growth rate 1990-2012} \hfill \null}
\label{tGenWhite} 
\small 
\vspace{-1.15cm}
\begin{tabular}{lcc} 
&  \textbf{Spatial model~\eqref{genSpat}} & \textbf{SAR White errors~\eqref{WhiteEq}} \\ 
\hline \\[-1.8ex]
     \vspace{-1.2cm} \\
 Constant & 0.174 (0.019)$^{***}$ & 0.177 (0.019)$^{***}$ \\ 
 Log of initial per capita income (US\$)& $-$0.033 (0.005)$^{***}$ & $-$0.033 (0.005)$^{***}$\\  
  Sea level rise (m/year) & 0.568 (0.235)$^{*}$ & 0.577 (0.244)$^{*}$ \\ 
  Sea level rise (m/year) - squared & $-$42.312 (30.441) & $-$43.675 (31.879) \\ 
  Coast distance (thousands km)& $-$0.004 (0.001)$^{***}$ & $-$0.004 (0.001)$^{***}$ \\ 
  Coast distance (thousands km) - sq. & 4185.800 (811.280)$^{***}$  & 4,347.300 (844.850)$^{***}$ \\ 
  Gov. expenditures per capita (bn. US\$) & $-$0.589 (0.572) & $-$0.590 (0.570) \\ 
    Tax income per capita (bn. US\$) & 3.219 (0.536)$^{***}$ & 3.330 (0.544)$^{***}$ \\ 
        $\rho$ (SAR) & 0.491 (0.053)$^{***}$ & 0.481 (0.054)$^{***}$ \\ 
  $\lambda$ (SEM) & $-$0.114 (0.078) & --- \\ 

\multicolumn{3}{c}{\textbf{Measures of agglomeration}}\\
   \vspace{-0.9cm} \\
 Population density (per thousand sq. miles) & $-$0.031 (0.116) & $-$0.035 (0.118)  \\
  Urban (dummy) & 0.001 (0.0003)$^{**}$ & 0.001 (0.0003)$^{**}$ \\ 
  Rural (dummy) & 0.0004 (0.0002) & 0.0003 (0.0003) \\ 
  \multicolumn{3}{c}{\textbf{Measures of religious adherence}}\\

     \vspace{-0.9cm} \\
    Catholics (percentage) & 0.0001 (0.00001)$^{***}$ & 0.0001 (0.00001)$^{***}$ \\ 
  Evangelical Protestants (percentage) & 0.0001 (0.00001)$^{***}$ & 0.0001 (0.00001)$^{***}$ \\ 
  Mainline Protestants (percentage) & 0.0001 (0.00002)$^{***}$ & 0.0001 (0.00002)$^{***}$ \\ 
    Religious diversity (Formula~\eqref{reldiv}) & 0.004 (0.001)$^{**}$ & 0.004 (0.001)$^{**}$ \\ 
  \multicolumn{3}{c}{\textbf{Other socioeconomic and environmental indicators}}\\
     \vspace{-0.9cm} \\
Education (percentage) & 0.0003 (0.00003)$^{***}$ & 0.0003 (0.00003)$^{***}$ \\ 
  Highway (dummy)& $-$0.0001 (0.0002) & $-$0.0001 (0.0002) \\ 
  Right to work laws (state level dummy) & 0.0010 (0.0003)$^{***}$ & 0.0010 (0.0003)$^{***}$ \\ 
  Nonwhites (percentage) & $-$0.0001 (0.00001)$^{***}$ & $-$0.0001 (0.00001)$^{***}$ \\ 
  Amenities (scale \citet{Amenities}) & $-$0.0002 (0.0001) & $-$0.0001 (0.0001) \\  
  \multicolumn{3}{c}{\textbf{Regional dummy variables}}\\
     \vspace{-0.9cm} \\
New England (dummy)& $-$0.003 (0.001)$^{***}$ & $-$0.003 (0.001)$^{***}$ \\ 
  Mideast (dummy)& $-$0.002 (0.001)$^{**}$ & $-$0.002 (0.001)$^{**}$ \\ 
  Great Lakes (dummy)& $-$0.003 (0.001)$^{**}$ & $-$0.003 (0.001)$^{**}$ \\ 
  Plains (dummy)& $-$0.003 (0.001)$^{**}$ & $-$0.003 (0.001)$^{**}$ \\ 
  Southeast (dummy)& $-$0.003 (0.001)$^{**}$ & $-$0.003 (0.001)$^{**}$ \\ 
  Southwest (dummy)& $-$0.002 (0.001)$^{*}$ & $-$0.002 (0.001)$^{*}$ \\ 
  Rocky Mountain (dummy)& $-$0.002 (0.001)$^{*}$ & $-$0.002 (0.001)$^{*}$ \\ 
  \hline
\multicolumn{3}{l}{\textit{Notes:} \hfill $^{*}$p$<$0.05; $^{**}$p$<$0.01; $^{***}$p$<$0.001, \hspace{0.5cm}Standard errors in brackets} \\ 
\end{tabular} 
\end{table} 

\linespread{1.6}
\begin{table}[H] \centering 
\caption{\small \textbf{\hspace{0.2cm} SAR White errors~\eqref{WhiteEq}}- Impact measures, \hfill \textit{1990-2012 }}
\label{tWhiteImpacts} 
\small 
\begin{tabular}{@{\extracolsep{-1pt}}lrrr} 
\vspace{-0.8cm} \\
\hline &  \textbf{Direct} &  \textbf{Indirect} &  \textbf{Total}\\ 
Sea level rise (m/year) &   0.6069 & 0.5045 & 1.1115 \\ 
Sea level rise (m/year) - squared & $-$45.9455 & $-$38.1962 & $-$84.1417 \\ 
Coast distance (thousands km) & $-$0.0046 & $-$0.0039 & $-$0.0085 \\ 
Coast distance (thousands km) - squared & 4,573.2200 & 3,801.8910 & 8,375.1110\\ 
Gov. expenditures per capita (billion US\$) & $-$0.6207 & $-$0.5160 & $-$1.1367 \\ 
Tax income per capita (billion US\$) & 3.5033 & 2.9124 & 6.4157\\  
\multicolumn{4}{c}{\textbf{Measures of agglomeration}}\\
Population density (rate per thousand square miles) &  $-$0.0367 &    $-$0.0305 &  $-$0.0672 \\
Urban (dummy) & 0.0009 & 0.0008 & 0.0017 \\ 
Rural (dummy)&  0.0004 & 0.0003 & 0.0006 \\ 
  \multicolumn{4}{c}{\textbf{Measures of religious adherence}}\\
Catholics (percentage) & 0.0001 & 0.0001 & 0.0001 \\ 
Evangelical Protestants (percentage) & 0.0001 & 0.0001 & 0.0001 \\ 
Mainline Protestants (percentage) & 0.0001 & 0.0001 & 0.0001 \\ 
Religious diversity (Formula~\eqref{reldiv}) & 0.0040 & 0.0034 & 0.0074 \\ 
  \multicolumn{4}{c}{\textbf{Other socioeconomic and environmental indicators}}\\
Education (percentage) & 0.0003  & 0.0003 & 0.0006 \\ 
Highway (dummy) & $-$0.0001  & $-$0.0001 & $-$0.0002 \\ 
Right to work laws (state level dummy) & 0.0010  & 0.0008 & 0.0019 \\ 
 Nonwhites (percentage) & $-$0.0001 & $-$0.0001 & $-$0.0002 \\ 
Amenities (scale \citet{Amenities})& $-$0.0002 & $-$0.0001 & $-$0.0003 \\ 
  \multicolumn{4}{c}{\textbf{Regional dummy variables}}\\
New England (dummy)& $-$0.0027 & $-$0.0022 & $-$0.0049 \\ 
Mideast (dummy)&$-$0.0024 & $-$0.0020  & $-$0.0043\\ 
Great Lakes (dummy)&  $-$0.0031 & $-$0.0026 & $-$0.0057 \\ 
Plains (dummy)&  $-$0.0028 & $-$0.0024 & $-$0.0052\\ 
Southeast (dummy)& $-$0.0026 & $-$0.0022 & $-$0.0048\\ 
Southwest (dummy)&  $-$0.0017 & $-$0.0015 & $-$0.0032 \\ 
Rocky Mountain (dummy)&  $-$0.0020 & $-$0.0017& $-$0.0037 \\ 
\hline \\[-1.8ex] 
\end{tabular} 
\end{table} 

\newpage

\linespread{1.575}
\begin{table}[H] \centering
\caption{\small \textit{\textbf{\hspace{0.2cm} Spatial autoregressive model~\eqref{SAR}},\hfill Growth rate between 1990-2012 } } 

\label{t30YrCompare} 
\small 
\vspace{-0.3cm}
\begin{tabular}{@{\extracolsep{-1pt}}lcc} 
\hline 
  \multicolumn{2}{r}{ \textbf{Whole periods of available data}} &  \textbf{SLR between} {$\boldsymbol{1979-2007}$} \\ 
Constant &  0.1849 (0.0074)$^{***}$ &  0.1842 (0.0074)$^{***}$ \\
Log of initial per capita income (US\$)  & $-$0.0333 (0.0049)$^{***}$ & $-$0.0333 (0.0049)$^{***}$\\ 
Sea level rise (m/year) & 0.5943 (0.2524)$^{*}$ & 0.5750 (0.3208)$^{\bullet}$\\ 
Sea level rise (m/year) - squared &  $-$44.4060 (33.7110) &  $-$69.2650 (58.3270)\\ 
Coast distance (thousands km) & $-$0.0045 (0.0012)$^{***}$  & $-$0.0052 (0.0011)$^{***}$\\ 
Coast distance (thousands km) - sq.  & 4535.1000 (690.0000)$^{***}$ & 4844.9000 (676.2900)$^{***}$ \\ 
Gov. expenditures per capita (bn. US\$)  & $-$0.5957 (0.4106) & $-$0.6337 (0.4106) \\ 
Tax income per capita (bn. US\$) & 3.3698 (0.3681)$^{***}$ &  3.3988 (0.3687)$^{***}$ \\ 
   $\rho$ (SAR) & 0.4583 (0.0206)$^{***}$ & 0.4610 (0.0205)$^{***}$ \\   
     \vspace{-0.9cm} \\
\multicolumn{3}{c}{\textbf{Measures of agglomeration}}\\
  \vspace{-0.9cm} \\
Population density (per thousand sq. miles) & $-$0.0213 (0.1303) & $-$0.0034 (0.1298) \\
Urban (dummy)  & 0.0009 (0.0003)$^{**}$ & 0.0009 (0.0003)$^{**}$\\ 
Rural (dummy)  & 0.0003 (0.0003) & 0.0003 (0.0003) \\ 
  \vspace{-0.9cm} \\
  \multicolumn{3}{c}{\textbf{Measures of religious adherence}}\\
  \vspace{-0.9cm} \\ 
Catholics (percentage) & 0.0001 (0.00001)$^{***}$ & 0.0001 (0.00001)$^{***}$ \\ 
Evangelical Protestants (percentage)  & 0.0001 (0.00001)$^{***}$& 0.0001 (0.00001)$^{***}$  \\ 
Mainline Protestants (percentage) & 0.0001 (0.00001)$^{***}$ & 0.0001 (0.00001)$^{***}$\\ 
Religious diversity (Formula~\eqref{reldiv})  & 0.0039 (0.0012)$^{***}$ & 0.0039 (0.0012)$^{***}$\\ 
  \vspace{-0.9cm} \\
  \multicolumn{3}{c}{\textbf{Other socioeconomic and environmental indicators}}\\
    \vspace{-0.9cm} \\ 
Education (percentage)& 0.0003 (0.00002)$^{***}$ & 0.0003 (0.00002)$^{***}$ \\ 
Highway (dummy)  & $-$0.0001 (0.0002) & $-$0.0001 (0.0002)  \\ 
Right to work laws (state level dummy) & 0.0010 (0.0003)$^{***}$ & 0.0010 (0.0003)$^{***}$\\ 
 Nonwhites (percentage)  & $-$0.0001 (0.00001)$^{***}$& $-$0.0001 (0.00001)$^{***}$ \\ 
Amenities (scale \citet{Amenities})& $-$0.0002 (0.0001)$^{*}$ & $-$0.0001 (0.0001)$^{\bullet}$ \\ 
  \vspace{-0.9cm} \\
  \multicolumn{3}{c}{\textbf{Regional dummy variables}}\\
    \vspace{-0.9cm} \\
New England (dummy)&  $-$0.0025 (0.0010)$^{**}$ &  $-$0.0027 (0.0010)$^{**}$\\ 
Mideast (dummy)&  $-$0.0023 (0.0008)$^{**}$ &  $-$0.0024 (0.0008)$^{**}$ \\ 
Great Lakes (dummy) & $-$0.0031 (0.0008)$^{***}$ & $-$0.0030 (0.0008)$^{***}$\\ 
Plains (dummy)&  $-$0.0028 (0.0009)$^{**}$ &  $-$0.0028 (0.0009)$^{**}$ \\ 
Southeast (dummy)&  $-$0.0026 (0.0007)$^{***}$ &  $-$0.0027 (0.0007)$^{***}$ \\ 
Southwest (dummy)&  $-$0.0017 (0.0007)$^{*}$ &  $-$0.0017 (0.0007)$^{*}$ \\ 
Rocky Mountain (dummy) & $-$0.0020 (0.0008)$^{**}$ & $-$0.0020 (0.0008)$^{*}$ \\ 
\hline \\
\vspace{-1.5cm} \\ 
\multicolumn{3}{l}{\textit{Notes:} \hfill $^{*}$p$<$0.05; $^{**}$p$<$0.01; $^{***}$p$<$0.001, \hspace{0.5cm}Standard errors in brackets} \\  
\end{tabular} 
\end{table} 
\newpage

\begin{table}[H] \centering
\caption{\small \textit{\textbf{SAR model~\eqref{SAR}} without government finances variables,\hspace{0.3cm} 1990-2012 }}
\label{tNoFinCompare} 
\small 
\begin{tabular}{lc} 
\vspace{-1cm} \\
\hline \\[-1.8ex] 
Constant &  0.1775 (0.0074)$^{***}$ \\
Log of initial per capita income (US\$)  & $-$0.0333 (0.0049)$^{***}$\\ 
Sea level rise (m/year) & 0.7199 (0.2564)$^{**}$\\ 
Sea level rise (m/year) - squared &  $-$51.3380 (34.2530)\\ 
Coast distance (thousands km)   & $-$0.0046 (0.0012)$^{***}$\\ 
Coast distance (thousands km) - squared & 4484.4000 (695.6300)$^{***}$ \\ 
   $\rho$ (SAR) & 0.4775 (0.0206)$^{***}$ \\   
\multicolumn{2}{c}{\textbf{Measures of agglomeration}}\\
Population density (rate per thousand square miles) &  $-$0.1609 (0.1305) \\
Urban (dummy)  &  0.0008 (0.0003)$^{*}$\\ 
Rural (dummy)  & 0.0005 (0.0003) $^{\bullet}$\\ 
  \multicolumn{2}{c}{\textbf{Measures of religious adherence}}\\
Catholics (percentage) & 0.0001 (0.00001)$^{***}$ \\ 
Evangelical Protestants (percentage)  & 0.0001 (0.00001)$^{***}$  \\ 
Mainline Protestants (percentage) & 0.0001 (0.00001)$^{***}$\\ 
Religious diversity (Formula~\eqref{reldiv})  & 0.0045 (0.0012)$^{***}$\\ 
  \multicolumn{2}{c}{\textbf{Other socioeconomic and environmental indicators}}\\
Education (percentage) & 0.0004 (0.00002)$^{***}$ \\ 
Highway (dummy)  &  $-$0.0002 (0.0002)  \\ 
Right to work laws (state level dummy) &  0.0015 (0.0003)$^{***}$\\ 
 Nonwhites (percentage) & $-$0.0001 (0.00001)$^{***}$ \\ 
Amenities (scale \citet{Amenities})& $-$0.0001 (0.0001)\\ 
  \multicolumn{2}{c}{\textbf{Regional dummy variables}}\\
New England (dummy)&   $-$0.0019 (0.0009)$^{*}$\\ 
Mideast (dummy)&  $-$0.0016 (0.0008)$^{*}$ \\ 
Great Lakes (dummy) & $-$0.0029 (0.0008)$^{***}$\\ 
Plains (dummy)&   $-$0.0029 (0.0009)$^{**}$ \\ 
Southeast (dummy)&   $-$0.0031 (0.0007)$^{***}$ \\ 
Southwest (dummy)&    $-$0.0016 (0.0007)$^{*}$ \\ 
Rocky Mountain (dummy)  & $-$0.0014 (0.0008)$^{\bullet}$ \\ 
\hline \\
\vspace{-1.5cm} \\ 
\multicolumn{2}{l}{\textit{Notes:} \hfill $^{*}$p$<$0.05; $^{**}$p$<$0.01; $^{***}$p$<$0.001, \hspace{0.5cm}Standard errors in brackets} \\  
\end{tabular} 
\end{table}

\end{document}